\documentclass[a4paper,12pt]{article}
\usepackage{subfigure}
\usepackage{amsmath,graphicx,amssymb}
\usepackage{multirow,color}
\usepackage[a4paper]{geometry}
\usepackage{tabularx}
\begin{document}
\renewcommand{\vec}[1]{\boldsymbol{\mathbf{#1}}}
\newcommand{\etal}{\textit{et al}}
\newcommand{\Cinf}{C_{\varepsilon,\infty}}
\newcommand{\kmax}{k_\text{max}}
\newcommand{\teP}{t_{\varepsilon\vert\Pi}}
\newcommand{\tE}{t_\varepsilon}
\renewcommand{\leq}{\leqslant}
\renewcommand{\geq}{\geqslant}

\newcommand{\smeasure}[1]{\mathscr{D}#1}
\newcommand{\measure}[1]{\smeasure{\vec{#1}}}
\newcommand{\pdf}[1]{P[\vec{#1}]}
\newcommand{\gen}[1]{Z[\vec{#1}]}
\newcommand{\ord}[1]{O\left(#1\right)}
\newcommand{\bal}{\begin{align}}
\newcommand{\eal}{\end{align}}
\newcommand{\set}[1]{\{#1\}}
\newcommand{\im}{\imath}

\newcommand{\half}{\frac{1}{2}}
\newcommand{\nono}{\nonumber}
\newcommand{\nuz}{\nu_{0}\,}
\newcommand{\vep}{\varepsilon}
\newcommand{\vepu}{\varepsilon_{U,L}}

\newcommand{\RL}{R_L}
\newcommand{\Rl}{R_{\lambda}}
\newcommand{\cdim}{C_{\vep}}
\newcommand{\Ceps}{C_{\vep}}
\newcommand{\cdiminf}{C_{\vep,\infty}}
\newcommand{\cpi}{C_{\Pi}}
\newcommand{\cpinf}{C_{\Pi,\infty}}

\newcommand{\beq}{\begin{equation}}
\newcommand{\eeq}{\end{equation}}
\newcommand{\bea}{\begin{eqnarray}}
\newcommand{\eea}{\end{eqnarray}}

\newcommand{\dd}{\partial}
\newcommand{\ddt}{\frac{\partial}{\partial t}}
\newcommand{\dr}{\frac{\partial}{\partial r}}

\newcommand{\xt}{(\mathbf{x},t)}
\newcommand{\kt}{(\mathbf{k},t)}
\newcommand{\pkt}{(\mathbf{k'},t')}
\newcommand{\kjt}{(\mathbf{k-j},t)}
\newcommand{\jpt}{(\mathbf{j-p},t)}
\newcommand{\jt}{(\mathbf{j},t)}
\newcommand{\lt}{(\mathbf{l},t)}
\newcommand{\kptp}{(\mathbf{k'},t')}
\newcommand{\kp}{k^{+}}
\newcommand{\ks}{(\mathbf{k},s)}
\newcommand{\kstar}{k_{\ast}}
\newcommand{\kb}{\ov{k}}
\newcommand{\kjs}{(\mathbf{k-j},s)}
\newcommand{\jps}{(\mathbf{j-p},s)}
\newcommand{\js}{(\mathbf{j},s)}
\newcommand{\ls}{(\mathbf{l},s)}
\newcommand{\ps}{(\mathbf{p},s)}

\newcommand{\up}{u^{+}}
\newcommand{\um}{u^{-}}
\newcommand{\ua}{u_\alpha}
\newcommand{\ub}{u_\beta}
\newcommand{\ug}{u_\gamma}
\newcommand{\ud}{u_\delta}
\newcommand{\us}{u_\sigma}
\newcommand{\ur}{u^{(r)}}
\newcommand{\tilur}{\tilde{u}^{(r)}}

\newcommand{\fa}{f_\alpha}
\newcommand{\fd}{f_\delta}
\newcommand{\Mp}{M_{\alpha\beta\gamma}^{+}(\mathbf{k})}
\newcommand{\Mm}{M_{\alpha\beta\gamma}^{-}(\mathbf{k})}
\newcommand{\Mk}{M_{\alpha\beta\gamma}(\mathbf{k})}
\newcommand{\Mmk}{M_{\alpha\beta\gamma}(-\mathbf{k})}

\newcommand{\Nk}{N_{\alpha\beta\gamma}(\mathbf{k})}
\newcommand{\Mpm}{M_{\alpha\beta\gamma}^{\pm}(\mathbf{k})}
\newcommand{\Npm}{N_{\alpha\beta\gamma}^{\pm}(\mathbf{k})}


\newcommand{\av}[1]{\left\langle #1 \right\rangle}
\newcommand{\cond}[1]{\left\langle #1 \right\rangle_{0}}
\newcommand{\cav}[1]{\left\langle #1 \right\rangle_{c}}
\newcommand{\rav}[1]{\left\langle #1 \right\rangle^{(r)}}
\newcommand{\ov}{\overline}


\title{Onset criteria for freely decaying isotropic turbulence}
\author{S. R. Yoffe\footnote{SUPA Department of
Physics, University of Strathclyde, John Anderson Building, 107
Rottenrow East. Glasgow G4 0NG.}\, and W. D. McComb\\
SUPA School of Physics and Astronomy,\\
Peter Guthrie Tait Road, \\
University of Edinburgh,\\
EDINBURGH EH9 3JZ.\\
Email: wdm@ph.ed.ac.uk}

\maketitle \thispagestyle{empty} \begin{abstract} From direct numerical
simulation (DNS) of turbulence decaying from specified initial
conditions for the range of initial Taylor-Reynolds numbers $2.58\leq
\Rl(0) \leq 358.6$, it was found that the shape of the iconic curve of
dimensionless dissipation versus Reynolds number depended strongly on
the choice of measurement time. For our preferred time, a composite
based on peak values in the dissipation and inertial transfer curves,
the result was virtually identical to the forced, stationary case. In
order to try varying the initial conditions, an additional run was
performed, using the data from a stationary, forced simulation with
$\Rl=335$ for the initial condition. The results of this suggested that
the time taken for energy to pass through the cascade was about one half
of an initial eddy turnover time. In the course of studying onset
criteria, we found that the exponent for the power-law decay of the
energy decreased with increasing Reynolds number and lay in the range
$1.35\leq n \leq 2.60$.

\end{abstract}

\newpage

\section{Introduction}

The present paper is the culmination of an investigation into the
dependence of the dimensionless dissipation rate $\Ceps$ on Reynolds number
in isotropic turbulence. Our approach over the last decade has been
based on analysis of the real-space energy balance equation (the
Karman-Howarth equation or KHE for short), with direct numerical
simulations providing validation and the evaluation of constants. Our
first work in this area was the reinterpretation of the Taylor
dissipation surrogate by McComb \etal  \cite{McComb10b} as a surrogate
for the inertial transfer, which becomes equal to the dissipation at
sufficiently large Reynolds numbers, corresponding to the development of
an inertial range. This work was for freely decaying turbulence. 

It is worth pointing out that the supplemental material for this paper
contained what was claimed to be an exact theoretical expression
relating the dissipation to the Reynolds number. This relied on a number
of physical arguments, adding up to a derivation that led to the form 
\beq
\Ceps(t_e) = \Cinf (t_e) + \frac{C^{decay}(t_e)}{\RL (t_e)},
\eeq
and this was readily fitted to our experimental results. Here $\RL$ is
the Reynolds number based on the integral scale $L$ (defined as
$L=\int_0^\infty \,f(r)\, dr = (3\pi/4E)\int dk\, E(k)/k$, where $f(r)$
is the longitudinal correlation function, $E(k)$ the energy spectrum, and
$E$ the total energy) while
 $C^{decay}(t_e)$ is a
coefficient derived from the second- and third-order structure functions
and the differential coefficient of the second-order structure function
with respect to time. A key feature is the fact that we had to evaluate
each term in this expression at some fiducial time $t=t_e$ during the
decay.

Unfortunately this derivation was not easy to understand and, in order to
make the theory more accessible, we resorted to an approximate method in
which we introduced asymptotic expansions of the structure
functions in powers of the inverted Reynolds number. This technique had
previously been used by Lundgren \cite{Lundgren02} to derive the
Kolmogorov two-thirds law. Logically the present paper should have been
the first outcome of this research, but our new method emphasised the
difficulties with time dependence in the decaying case. So we turned
instead to the derivation of an asymptotic theory of the dissipation
rate in forced turbulence, and its verification using DNS: see McComb
\etal \cite{McComb15a}. The result was similar to that for the decaying
case, but of course without the time dependences, thus:
\beq
\Ceps = \Cinf + \frac{C}{\RL},
\eeq
where $\Cinf$ depends on the third-order structure function, while $C$
depends on both  the second- and third-order structure functions. In
principal we could include higher-order terms: the second-order term in
the asymptotic expansions was needed when the work was extended to free
decay of homogeneous magnetohydrodynamic turbulence by Linkmann \etal
\cite{Linkmann15a}. However, several numerical tests suggested that the
dependence on the Reynolds number is as indicated in (2), thus lending
support to the idea that this is actually an exact result.

The extension of our new theory to freely decaying isotropic turbulence
was done later by McComb and Fairhurst \cite{McComb18a} and leads to two
principal results. These are:
\begin{description}
\item[A] That the time-derivative in the KHE which was neglected in the
derivation of the Kolmogorov `4/5' law
\cite{Kolmogorov41A,Kolmogorov41B} cannot be so neglected solely on the basis
of a restriction to certain scales or to large Reynolds numbers.
\item[B] That the neglect of this term can be quantified by a comparison
of the free decay and forced forms of the asymptotic dimensionless dissipation
rate, provided only that we know the correct value $t=t_e$ to choose as a
fiduciary time. 
\end{description}

The importance of this is that Kolmogorov \cite{Kolmogorov41A}
introduced the concept of \emph{local stationarity} (as an aspect of
local isotropy) and that this was further described by Batchelor
\cite{Batchelor53} as \emph{local equilibrium}. These were presented as
general properties; and are not quite the same as arguments that the time
derivative may be neglected as an approximation on one set of scales in
comparison to another. Yet in practice one suspects that the two concepts
are often blurred. Accordingly, it is helpful to have the general
concepts ruled out as exact, rigorous properties, But the need to
quantify local stationarity as an approximation provides an additional
motivation (were one needed) to establish to what extent the asymptotic
dimensionless dissipation rate depends on the decay time when the
measurement is made, and that is the subject of this paper.

The paper is organised into the following sections:

Section 2 provides a short review of the study of the dependence of the
dimensionless dissipation rate on Reynolds number with particular
emphasis on the topics that are most relevant to the present work.

Section 3 discusses the nature of the problem of deciding when a
decaying simulation is fully evolved, in comparison with the problem of
doing this for a forced simulation, and proposes various possible onset
criteria.

Section 4 presents details of the DNS and also gives the results for
various statistical quantities as a function of both elapsed time and
Reynolds number in order that the quality of the simulations can be
assessed. This allows an assessment of criteria for choosing an
evolved time based on the behaviour of the dissipation rate, the
inertial transfer rate and the skewness factor.

Section 5 proposes a composite criterion, based the evolution in time of
both the dissipation and the inertial transfer rates.

Section 6 gives the effect of individual choices of evolved time on the
measurements of the dissipation rate.

Section 7 discusses the results, and puts forward suggestions for future
work. 

Appendix A is a brief introduction to power-law decay and compares our
results to the field in general

Appendix B explores the alternative determination of the energy exponent by
measuring the Taylor microscale.

\section{The dimensionless dissipation rate}

In recent years, there has been great interest in the Reynolds number
dependence of the dissipation rate in homogeneous, isotropic turbulence
(HIT). Apart from its intrinsic fundamental signifance, it is a key
factor in the free decay of a turbulent fluid. Much of this work has
been based on Taylor's expression for the dissipation rate, thus:
\beq
\varepsilon = \cdim(\RL) U^3/L,
\label{taylordiss}
\eeq
which was put forward in 1935 by Taylor \cite{Taylor35} on the basis of
dimensional arguments. Here $U$ is the rms velocity of the fluid and $L$
is the integral length scale. Note that we have explicitly indicated the
dependence of $\cdim$ on the Reynolds number. Nowadays it is more usual
to rearrange this as
\beq
\cdim(\RL) = \varepsilon L/ U^3,
\label{dimdiss}
\eeq
so that $\cdim(\RL)$ may be interpreted as the dimensionless dissipation
rate, rather than the Taylor prefactor. 
As early as 1953, Batchelor \cite{Batchelor71} (in the first edition of
this book) presented evidence to suggest that $\cdim$ tends to a
constant $\cdiminf$ with increasing Reynolds number. Later Sreenivasan
\cite{Sreenivasan84} established, from a survey of investigations into
grid-generated turbulence, that $\cdim$ became constant for
Taylor-Reynolds numbers greater than about 50. This independence of
viscosity is sometimes referred to as the \emph{dissipation anomaly}. 

For a Newtonian fluid, the dissipation rate
is formally defined in terms of the kinematic viscosity $\nu$, thus:
\beq
\hat{\varepsilon} = \frac{\nu}{2}\left(\frac{\dd u_i}{\dd x_j} + \frac{\dd
u_j}{\dd x_i}\right)^2.
\label{dissdefn}
\eeq
As $\mathbf{u}\xt$ is a random variable with zero mean, it follows that
$\hat{\varepsilon}$ is the \emph{instantaneous} dissipation rate,
and is also a random variable. For a turbulent flow we introduce the
\emph{mean} dissipation rate, as: \beq
\varepsilon = \av{\frac{\nu}{2}\left(\frac{\dd u_i}{\dd x_j} + \frac{\dd
u_j}{\dd x_i}\right)^2},
\label{mean-dissdefn}
\eeq
where the angle brackets $\av{\dots}$ denote the operation of taking an
average. To avoid any possible confusion, it should be noted that we
will use the \emph{mean} dissipation rate, along with the other averaged
quantities, in our description of fluid turbulence. Accordingly, the
concept of intermittency, which is related to the behaviour of single
realizations, and is just one aspect of the phase-dependent behaviour of
the velocity field, has no relevance to our discussion here. Here we
begin by reviewing the relevant literature and summarising recent work
on the behaviour of the dimensionless dissipation as the Reynolds number
is increased. 

Sreenivasan \cite{Sreenivasan84} had concluded that results for
square-mesh grids suggested that the dimensionless dissipation rate
became independent of Reynolds number for $\Rl > 50$. He noted that
there was a 
marked variation at lower Reynolds numbers but found this unsurprising.
Where the inertia forces could be neglected, one might expect a
relationship of the form \beq
\cdim = 15(\pi/2)^{1/2}/\Rl.
\eeq
This arose from the well known expression $\vep = 15\nu
U^2/\lambda^2$, taken in conjunction with the approximation $L/\lambda
\simeq (\pi/2)^{1/2}$. He also remarked upon the lack of clear evidence
for other flow configurations, but speculated that the asymptotic value
of the dimensionless dissipation could depend on the initial conditions.

In an update, in 1998, Sreenivasan \cite{Sreenivasan98} presented
data from four investigations using DNS. These were by Jim\'enez, Wray,
Saffman and Rogallo \cite{Jimenez93},  Yeung and Zhou \cite{Yeung97} and
Cao, Chen and Doolen \cite{Cao99}, all studying forced turbulence, and
by Wang, Chen, Brasseur and Wyngaard \cite{Wang96}, who studied both
decaying and forced turbulence.  He concluded that
$C_{\varepsilon,\infty} \sim O(1)$, with $C_{\varepsilon,\infty}$
sensitive to both the initial conditions and the method of forcing. Wang
\etal\ \cite{Wang96} found $\Cinf \sim 0.62$ (decay) and 0.42 --
0.49 (forced). 

In considering the scatter of results, Sreenivasan highlighted the
relevance of large-scale resolution, noting that in the forced
simulations the integral scale and the box size were about the same.
More recently, Burattini, Lavoie and Antonia \cite{Burattini05}
presented an experimental investigation of $\Ceps$ for several different
experimental setups including grid turbulence (both active and passive
grids) and a variety of bluff body wakes. Their figures 1 and 2
summarize their results, which show a spread of  $\Cinf$ values from 0.5
-- 2.5 for $\Rl > 50$. These authors drew attention to the difficulties
of measuring the dissipation rate, remarking that grid turbulence has
the advantage that it can be determined from the decay rate. They noted
that the asymptotic value of the dimensionless dissipation not only
varied from one flow type to another, but also could depend on details
of the body producing the turbulence. 

\begin{figure}[tbp]
 \begin{center}
  \includegraphics[width=0.6\textwidth]{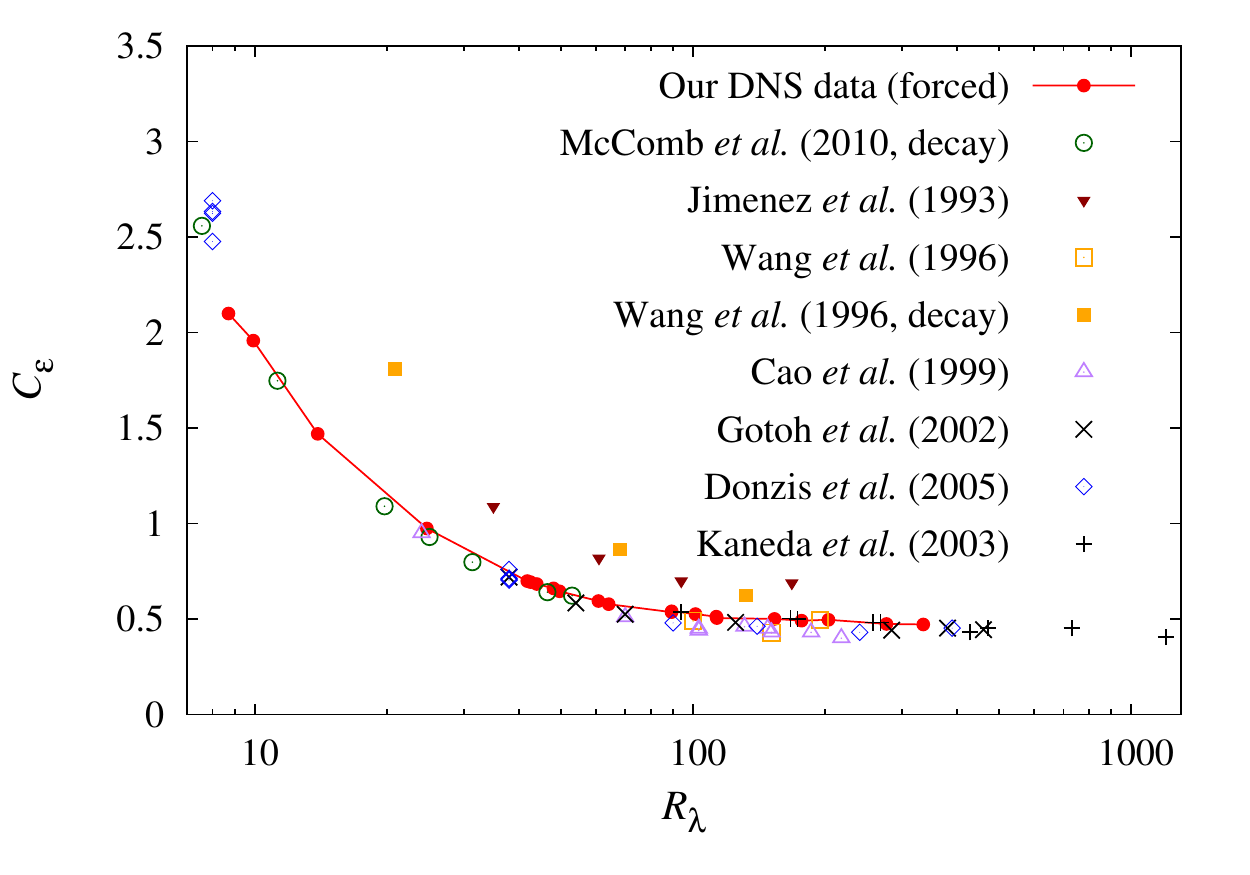}
 \end{center}
 \caption{A review of the data obtained for $\Ceps$ as a function of 
 Taylor-Reynolds number, $R_\lambda$, from numerical simulations. Data sets 
 from decaying turbulence are indicated in the figure legend. Along with 
 data obtained the present work, and from McComb \etal\ \cite{McComb10b}, we 
 also plot results from Jim\'enez \etal\ \cite{Jimenez93}, Wang \etal\ 
 \cite{Wang96}, Cao \etal\ \cite{Cao99}, Gotoh \etal\ \cite{Gotoh02}, 
 Donzis \etal\ \cite{Donzis05} and Kaneda \etal\ \cite{Kaneda03}.}
 \label{fig:Ceps_review}
\end{figure}

However, for forced DNS at least, there seemed to be a general trend to
an asymptotic value of slightly less than $0.5$. In an adaptation of the
figure in \cite{Sreenivasan98}, Burattini \etal \cite{Burattini05}
presented additional results for forced DNS from Gotoh, Fukuyama and
Nakano \cite{Gotoh02} and from Kaneda, Ishihara, Yokokawa, Itakura and
Uno \cite{Kaneda03}. Here we also present an adaptation of this figure,
as Fig. \ref{fig:Ceps_review}, where we have added results for DNS of
free decay from McComb, Berera, Salewski and Yoffe \cite{McComb10b},
along with some new results from the present investigation. For forced
DNS we have also added results from Donzis, Sreenivasan and Yeung
\cite{Donzis05}.

Measurements of $\Ceps$ were made by Pearson, Krogstad and van der Water
\cite{Pearson02} in various shear flows (including several forms of
grid-generated turbulence). From our present point of view,  their
conclusion that for $\Rl \geq 300$  a value of $\Ceps \sim 0.5$ seemed
to be a good universal approximation for flow regions free of strong
mean shear, is of particular interest. Later Pearson and co-workers
\cite{Pearson04a} used DNS to study stationary HIT, up to a
Taylor-Reynolds number of about 220. They found that the dimensionless
dissipation rate slowly approached a value of $\Ceps \sim 0.5$, although
we should note that their simulation was slightly compressible.

A rather more extensive investigation than most was carried out by Bos,
Shao and Bertoglio \cite{Bos07}, who used a variety of methods to study
both decaying and forced turbulence. Pseudospectral DNSs were carried
out on $64^3$ -- $256^3$ lattices, with a maximum value of $R_\lambda
\sim 100$ for their forced runs. They also used  large-eddy simulation
(LES) and the well-known single-time model closure EDQNM, with
$R_\lambda$ up to $\sim 2000$.

For the simulations of free decay, they tried four different initial
spectra and concluded that, after a short transient, the different
simulations all led to the same value of the dimensionless dissipation.
In the case of the forced simulations, they tried three different
forcing schemes, and found that all three ended up with values of
$\Ceps$ which oscillated about a constant value of about $\Ceps = 0.6$.
Again the only difference was in the initial transient and, in this
case, the size of the oscillations about the mean value.

The techniques of LES and EDQNM were used to extend the investigation to
much higher Reynolds numbers. The clear result was a significant
difference between the decaying and forced cases. As mentioned above
they found $\Cinf \sim 0.6$ for the forced case while for free decay the
asymptotic dimensionless dissipation took the much larger value of $\Cinf 
\sim 1$. 

Bos \etal\ discussed their results in the context of other work. At
low Reynolds numbers they found that their results tended to the form
$a\RL^{-1}$, as predicted by Sreenivasan (see above), with $a \sim 20$.
They noted reasonable agreement with the result of Mydlarski and Warhaft
\cite{Mydlarski96}, who found $\Cinf \sim 0.9$ for grid-generated
turbulence, but remarked upon the disagreement with Pearson \etal\ 
\cite{Pearson02}, who obtained a much lower value, and speculated about
the difficulty of measuring the integral scale and also the possibility
of finite-size effects. A particularly interesting feature of this work
is their conclusion on phenomenological  grounds that the value of
$\Cinf$ \emph{should} be larger for free decay than for forced HIT. We
shall return to this later in the context of our own work in the present
paper. At this point we mention our earlier work on the dimensionless
dissipation in freely decaying turbulence.

McComb, Berera, Salewski and Yoffe \cite{McComb10b} made a numerical
investigation of energy dissipation and transfer rates in  decaying
isotropic turbulence with  $R_\lambda(t_e) < 60$.
They tried three different initial spectra, with both $k^2$ and $k^4$ 
low-wavenumber behaviour, and measured the dimensionless dissipation
rate $\Ceps$ using the time at which the peak of the dissipation
rate occurred as their measurement time. The results were comparable
to forced simulations, with $\Cinf \sim 0.5$. They also compared the
dissipation rate with the transfer rate and the dissipation  surrogate
$U^3/L$ for each of the three initial spectra, and showed that $U^3/L$ was a 
better surrogate for transfer than the dissipation rate. A plot of
the ratio $\varepsilon/\varepsilon_T$ (again, measured at the peak of
the dissipation rate, $t_e$) for the three spectra, showed the approach
to $\varepsilon_T = \varepsilon - \Delta$, where $\Delta$ is due to the
finite cascade time and decaying nature of the turbulence.

Since then we have developed theoretical arguments to support our
conclusion that Taylor's expression is actually a surrogate for the
inertial transfer rate (rather than for the dissipation), and have found
that results from the present DNS support this view. We do not claim
that this interpretation is entirely new: some commentators seem to have
taken the same view (for instance, see the books by Batchelor
\cite{Batchelor71}, Tennekes and Lumley \cite{Tennekes72}, Davidson
\cite{Davidson04} and Sagaut and Cambon \cite{Sagaut08}). However, it
certainly runs counter to the view which underpins many investigations
in laboratory and environmental flows, along with more recent work on
DNS. 

We should also mention the comprehensive study of decaying turbulence by
Vassilicos and co-workers
\cite{Seoud07}-\nocite{Mazellier08}\nocite{Valente11a}\cite{Valente12b},
who used a variety of fractal grids, active grids and regular grids.
This work underlines the need to understand better what is truly
universal in turbulence. Their various values for $\Cinf$ are given,
along with those of others, in Table \ref{tbl:Cinf_review}. Note that
for fractal grids they found that the dimensionless dissipation rate does
not seem to have a finite asymptotic value as the Reynolds number tends
to infinity, whereas for regular grids they favour a value of $\Cinf$
about unity. In contrast, Krogstad and Davidson \cite{Krogstad11} found
that $\Ceps \sim 0.6$ for both multiscale and conventional grids at a
Taylor-Reynolds number of about 70. It should be noted that these
authors plot (in effect) $3\Ceps /2$ against downstream distance in
their figure 11. 

Lastly, for completeness we note the investigation of the relationship
between \emph{self-preservation} and the dependence of the dissipation
rate on Reynolds number in two kinds of decaying turbulence by Djenidi
\etal \cite{Djenidi17}, as the most recent work of this group.

 \begin{table}
  \begin{center}
   \begin{tabularx}{\textwidth}{lllX}
    Reference & $\Cinf$ & $R_\lambda$ & Comments \\
    \hline
    Sreenivasan \cite{Sreenivasan84} & $\sim 1.0$ & 5--500 & Experimental \\
    Jim\'enez \etal\ \cite{Jimenez93} & $\sim 0.65$ & 35--168 & DNS (forced)\\
    Wang \etal\ \cite{Wang96} & $\sim 0.42$--0.49 & 100--195 & DNS (forced) \\
    & $\sim 0.62$ & 21--132 & DNS (decay) \\
    Yeung and Zhou \cite{Yeung97} & $\sim 0.4$ & 38--240 & DNS (forced)\\
    Sreenivasan \cite{Sreenivasan98} & $\sim 0.5$ & 20--250 & DNS from 
    \cite{Wang96,Cao99,Yeung97} (forced) \\
    & $\sim 0.78$ & 35--195 & DNS from \cite{Jimenez93,Wang96} (forced, decay) \\
    Cao \etal\ \cite{Cao99} & 0.39--0.45 & 24--218 & DNS (forced)\\
    Pearson \etal\ \cite{Pearson02} & 0.4--0.6 & 50--1200 & Experimental\\
    Gotoh \etal\ \cite{Gotoh02} & $\sim 0.5$ & 38--460 & DNS (forced) \\
    Kaneda \etal\ \cite{Kaneda03} & $0.4 \sim 0.5$ & 94--1201 & DNS (forced) \\
    Pearson \etal\ \cite{Pearson04a} & $\simeq 0.5$ & 50--1200 & DNS 
    (forced), compressible \\
    Donzis \etal\ \cite{Donzis05} & $\sim 0.5$ & 10--400 & DNS (forced) \\
    Burattini \etal\ \cite{Burattini05} & 0.5--2.5 & 50--1100 & 
    Experimental \\
    Bos \etal\ \cite{Bos07} & $\sim 0.6$  & 0--2000 & DNS, LES and EDQNM
    (forced) \\
    & $\sim 1$ & 0--2000 & DNS, LES and EDQNM (decay)\\
    McComb \etal\ \cite{McComb10b} & $\sim 0.5$ & 3--60 & DNS (decay) \\
    Krogstad and Davidson \cite{Krogstad10} & $\sim 1$ &  & 
    Experimental (regular grid) \\  
    Valente and Vassilicos \cite{Valente11a} & $\sim 0.5$ & 100-570 & 
    Experimental (various grids)\\
    Valente and Vassilicos \cite{Valente12b} & 0.9--1.0 & & 
    Experimental (regular grid)\\
   \end{tabularx}
  \end{center}
  \caption{Some representative values for the asymptotic dimensional
dissipation rate $\Cinf$ from the literature. Bos \etal\ 
\cite{Bos07} obtained $R_\lambda \leq 2000$ using LES, and $R_\lambda \leq 
100$ using DNS.}
  \label{tbl:Cinf_review}
 \end{table}

\section{The nature of the problem}
Free decay of isotropic turbulence is the principal benchmark problem
for both phenomenological and theoretical treatments of the turbulence
problem. Essentially it is believed that the turbulence kinetic energy
decays as a power law, that is: $U^2 \sim t^{-n}$, where the exponent
$n$ has to be determined. But, more than seven decades after the study
of this problem began, there is an absence of agreement or even much
consensus on the value of this exponent. 

When formulated as a problem in mathematical physics, and realized in
practice in a direct numerical simulation (DNS), it is an initial value
problem where the initial velocity field is arbitrarily chosen to have a
Gaussian distribution and a specified form of energy spectrum. As the
system evolves in time (or is iterated forward in time), one expects to
achieve a velocity field which is determined by the Navier-Stokes
equations. In this respect it is like the problem of forced
isotropic turbulence. But in other respects it is not. In principle (and
in practice) we can run a forced simulation forward in time until we are
sure that we are observing Navier-Stokes turbulence. The end state is
stationary turbulence which fluctuates about a fixed mean value of the
energy, with the fluctuating mean dissipation rate lagging behind,
reflecting the time needed for energy to pass through the cascade
\cite{McComb01a}. It should perhaps be emphasised that these are not
turbulent fluctuations, but are analogous to the fluctuations in total
energy of the canonical ensemble of statistical physics. As in that
case, the relative magnitude of the fluctuations should decrease as the
system size is increased.

However, in the case of free decay, the energy is already decaying, even
before turbulence has been established. This raises a question which, as
far as we know, has not been formally addressed: at what time in the
decay process, $t=t_e$ (say), can the turbulence be said to be fully
evolved? And what exactly do we mean by `fully evolved'? 

Inevitably the situation is complicated by the historical fact that the
problem has been mostly studied experimentally in the form of
grid-generated or grid turbulence. Thus current assessments tend to
consist of a mix of results from laboratory experiments and increasingly
from DNS (i.e. numerical experiments). Hence, the problem in making
comparisons --- that is, of `like with like' ---  lies in the detailed
specification of the initial conditions. 

Since the late 1950s, the formulation of the theoretical
problem has been based on an analogy with statistical physics. The
initial field is taken to be Gaussian and to have a spectrum which is
peaked near the origin. Development with time then involves the
nonlinear coupling process transferring energy to ever higher
wavenumbers, until the process is limited by viscosity. Thus, providing
only that the initial spectrum satisfies some quite weak conditions, the
resulting turbulence should be universal. The extension to DNS can and
should follow this route, to the extent permitted by the restrictions
imposed by the computational process.

In the case of grid turbulence, the situation is very much more
complicated. To begin with, strictly it involves free decay of a
stationary flow in space rather than time, so that a Galilean
transformation is required to interpret the problem as one of time
dependence. Then, there can be many choices of initial conditions,
depending on the type and solidity of grid. There can also be other
complicating factors to do with the design of the wind tunnel. In short
we may have little idea of whether any given grid turbulence experiment
satisfies the requirements of the theoretical/computational formulation.
Hence it behoves us to be  cautious about drawing parallels between DNS
and any particular laboratory experiment. Indeed, the need to reconcile
DNS simulations with grid turbulence is a developing field of research:
see the recent work by Meldi and Sagaut \cite{Meldi18}, which assesses
the effect of perturbations on the energy spectrum.

At the same time, it is not surprising that the field should be so
controversial and confused. But, while we do not attempt to offer any
remedy for that here, we do think that it emphasises the need to have a
clear understanding of at least the theoretical/numerical formulation
and assisting in this is our purpose in the present paper.

\subsection{Possible criteria for an evolved decay time $t=t_e$}

Naturally if one is concerned with measuring the exponent $n$ in the
power law for the decay of the total energy, then the onset criterion
should be based on the occurrence of power-law behaviour. Even then
there are two ways of doing this. We could use a log-log plot of energy
against decay time, or we can rely on the well known result that the
Taylor microscale $\lambda$ is proportional to the square root of the
decay time when the decay follows a power law, irrespective of the value
of the exponent (see, for example, Section 7.5.1 of reference
\cite{McComb14a}. This criterion has also been used in the context of
measurements of $\Ceps$ by Wang \etal \cite{Wang96} and Bos \etal
\cite{Bos07}. 

However, significant variation in the dimensionless dissipation rate
occurs at very low Taylor-Reynolds numbers and this raises the
possiblity that our choice of evolution time $t_e$ should be based in
some way on the evolution of the dissipation rate, as was done by
Fukayama \etal\ \cite{Fukayama00} and by McComb \etal \cite{McComb10b}.
Also, ideally we would like the time $t_e$ to be as small as possible,
since the system is decaying; and the longer we wait the smaller the
time-series that we have to work with.

There are at least four possible candidates for the choice of a
criterion to determine the evolved time. These are as follows:
\begin{enumerate}
\item The onset of power-law decay of the total turbulence energy;
\item The onset of $t^{1/2}$ scaling of the Taylor microscale $\lambda$;
\item The occurrence of peak dissipation and/or transfer rate;
\item The occurrence of a peak in the skewness of the longitudinal
velocity derivative when plotted against time.
\end{enumerate}
It should be noted that the first two criteria are essentially variants
of the same thing. Nevertheless,  we shall treat each of these four
potential criteria separately.

\subsection{Development of nonlinear energy transfer}

When considering whether the flow is fully developed, we know that
the fluid motion is initially random but not turbulent. The main
characteristic of turbulence is the growth of inertial transfer of
kinetic energy by means of nonlinear coupling. Let us denote the
inertial/nonlinear transfer rate by
\begin{equation}
  \varepsilon_T(t) = \text{max}\ \Pi(k,t) \ ,\qquad \Pi(k,t) = \int_k^\infty 
  dk\ T(k,t) \ ,
\end{equation}
where $\Pi(k,t)$ is the transport power spectrum and $T(k,t)$ the
transfer spectrum which may be expressed in terms of the triple moment,
in the usual way \cite{McComb14a}. Hence $\varepsilon_T$ 
represents the maximum rate of energy  transfer through wavenumbers.

It is instructive to make a comparison between decaying and forced
turbulence. We begin with the latter.
\begin{description}
\item[Forced turbulence:] \
 \begin{enumerate}
  \item At low Reynolds numbers, energy is also dissipated from low 
  wavenumbers, so not all energy passes through the cascade. Therefore, 
  we may expect to find $\varepsilon_T < \varepsilon$.
  \item At higher Reynolds numbers, dissipation from low wavenumbers becomes 
  negligible, and all energy dissipated must travel through the cascade.
In this case we expect to find $\varepsilon_T = \varepsilon$.
 \end{enumerate}
Also, since the flow is forced and in a steady state, we must have the input 
 $\varepsilon_W = \varepsilon$.
\item[Decaying turbulence:] \
\begin{enumerate}
\item Even at high Reynolds numbers, we expect that 
 $\varepsilon_T(t) = \varepsilon(t+\Delta t) < \varepsilon(t)$, where the 
 equality implies that energy transferred at time $t$ will be dissipated at 
 a later time, $t+\Delta t$, and the inequality is due to the fact that the 
 system is decaying.
\item McComb \etal \cite{McComb10b} showed that 
 $U^3/L$ is a better surrogate for $\varepsilon_T$ than $\varepsilon$. In 
 this case,
 \begin{equation}
  \Ceps = \frac{\varepsilon L}{U^3} \sim \frac{\varepsilon}{\varepsilon_T} \ 
  ,
 \end{equation}
 and the approach of $\Ceps$ to a plateau can be seen as the development of 
 an inertial range.
\item For the case of decaying turbulence, $\varepsilon(t) > 
 \varepsilon_T(t)$ and so the measured plateau will satisfy 
 $\Cinf^\text{decay} \geq \Cinf^\text{forced}$. See also the recent work
on this particular point by McComb and Fairhurst \cite{McComb18a}.
\end{enumerate}
\end{description}

If one starts with energy concentrated in the low wavenumbers; then, 
during the transition period, when the turbulence is developing, one
can  measure $\varepsilon_T(t) > \varepsilon(t)$ as the energy is
redistributed  among modes. This is an indication that the turbulence is
not fully developed.   It is important that the measurement time $t=t_e$
should be taken late enough in the decay in order to ensure that
turbulence has been properly established, but early enough for larger
Reynolds numbers to be explored.

\section{Numerical simulations}

We used a pseudospectral DNS, with full dealiasing performed by truncation 
of the velocity field according to the two-thirds rule. Time advancement for 
the viscous term was performed exactly using an integrating factor, while 
the nonlinear term used Heun's method (second-order predictor-corrector).  
In general, each run was started from a Gaussian-distributed random field with an 
initial energy spectrum (which behaves as $k^4$ for the low-$k$ modes),
\begin{equation}
 \label{eq:intial_spec}
 E(k,0) = c (k/k_0)^4 \exp\left[ -(k/k_0)^2 \right] \ , \qquad c = 0.266,\ 
 k_0 = 3.536 \ .
\end{equation}
The one exception was a simulation started from an evolved
$1024^3$ field.

For each Reynolds number studied, the only initial condition changed was
the value assigned to the (kinematic) viscosity, $\nu_0$. An ensemble
was generated by starting the simulations from different random initial
field  configurations (which all have the same $E(k,0)$ defined above).
This  ensemble, together with shell averaging, was used to calculate
statistics. The ensemble size for each Reynolds number discussed
in this work was 10, with the exception of decay from an evolved
$1024^3$ field, for which  only an ensemble of five realisations could
be obtained. Simulations were  run using lattices of size $64^3,\
128^3,\ 256^3,\ 512^3\ \text{and}\  1024^3$, with corresponding initial
Reynolds numbers ranging from  $R_\lambda(0) = 2.58$ up to $358.6$.
Since the simulations are decaying,  measurements are made at
progressively lower Reynolds number. The smallest  resolved wavenumber
was $k_\text{min} = 2\pi/L_\text{box} = 1$ in all  simulations, while
the maximum wavenumber always satisfied $k_\text{max}  \eta > 1.0$,
where $\eta$ is the Kolmogorov dissipation lengthscale\footnote{For
 quantitative consideration of the resolution of the dissipation length
scales, see Fig. 2 in \cite{McComb01a} and Fig. 5.2 of \cite{Yoffe12}}. 
The integral
scale, $L$, was found to lie between $0.35 L_{\text{box}}$ for our
lowest Reynolds number and $0.18 L_{\text{box}}$ for our largest.
Spectral quantities have been shell averaged using the smallest 
space-filling shell width, $\Delta k = 1$. A summary of these
simulations is given in table \ref{tbl:summary_decay_sims}.

\begin{table}[tbp!]
\begin{center}
\begin{tabular}{r|lll|l|ll}
$N$ & $\nu_0$ & $R_L(0)$ & $R_\lambda(0)$ & $t_{\text{max}}$ &
$R_L(t_{\text{max}})$ & $R_\lambda(t_{\text{max}})$ \\
\hline
\hline
128  & 0.1     & 3.236  & 2.582  & 12.9$\tau(0)$  &  0.1702  &  0.1512 \\
128  & 0.07    & 4.623  & 3.688  & 12.9$\tau(0)$  &  0.3861  &  0.3217 \\
128  & 0.05    & 6.473  & 5.164  & 12.9$\tau(0)$  &  0.7795  &  0.6209 \\
128  & 0.03    & 10.79  & 8.606  & 12.9$\tau(0)$  &  2.132   &  1.615  \\
128  & 0.02    & 16.18  & 12.91  & 12.9$\tau(0)$  &  4.284   &  3.048  \\
128  & 0.01    & 32.36  & 25.82  & 64.4$\tau(0)$  &  4.477   &  3.705  \\
128  & 0.007   & 46.23  & 36.88  & 64.4$\tau(0)$  &  6.920   &  5.303  \\
128  & 0.005   & 64.73  & 51.64  & 64.4$\tau(0)$  &  9.679   &  6.845  \\
256  & 0.0025  & 129.5  & 103.3  & 64.4$\tau(0)$  &  17.55   &  10.65  \\
256  & 0.0018  & 179.8  & 143.4  & 64.4$\tau(0)$  &  22.76   &  12.92  \\
512  & 0.00072 & 449.5  & 358.6  & 51.7$\tau(0)$  &  63.76   &  26.42  \\
\hline
1024 & 0.0002  & 3828.2 & 353.7  & 3.19$\tau(0)$  &  1742.2  &  182.9 \\
\end{tabular}
\caption{Summary of the simulations and 
their initial parameters. With the exception of the $1024^3$ run, all runs 
started from a Gaussian random-field with an initial energy spectrum which 
goes as $k^4$ at low $k$ and an ensemble of 10 was created by using 
different seeds to the initial field generation. The $1024^3$ run instead 
used an ensemble of 5 simulations which started from evolved fields obtained 
from a stationary simulation. The total simulation time is given by 
$t_\text{max}$ in terms of the initial value of the large-eddy turnover
time $\tau(0)=L(0)/U(0)$.}
\label{tbl:summary_decay_sims}
\end{center}
\end{table}

Our numerical code has been benchmarked using established test problems
such  as the Taylor-Green vortex \cite{Taylor37} and our results are in
agreement with those of Brachet \etal \cite{Brachet83}. Comparison has
also been made to data obtained using the  freely-available
pseudospectral code \emph{hit3d}\footnote{\emph{hit3d} is  available
from \texttt{http://code.google.com/p/hit3d/}}, with good agreement
being found. Since an important constraint of this work is the condition
of  isotropy, we have verified that the isotropy spectra measured for
the DNS  steady-state ensembles do not exhibit significant deviation
from an isotropic system. This was done by comparing the average energy
in the  directions of two randomly-orientated unit vectors for each
wavevector. Full details of benchmarking, statistics and the error
analysis will be found in \cite{McComb14b,McComb15a} and in the thesis
by Yoffe \cite{Yoffe12}. In addition, energy and transfer spectra for some
of these simulations, as well as integral parameters, may be found in
Section 4.1, pages 102--105 of the same thesis. We note also that simulation
data from this investigation is available to download.

\subsection{Resolution of the integral scales during the simulations}

Once a simulation is started, the total energy simply begins to decay.
But energy spectra show a more complicated behaviour and evolve from the
initial chosen shape as the energy spreads out to increasing wavenumber.
Similarly the transfer spectrum evolves from an initial value of zero.
These facts are well known and we do not report such results in detail
here. Instead, we concentrate on those results which are most relevant
to the rest of the paper.

\begin{figure}[tbp]
 \begin{center}
  \includegraphics[width=0.6\textwidth]{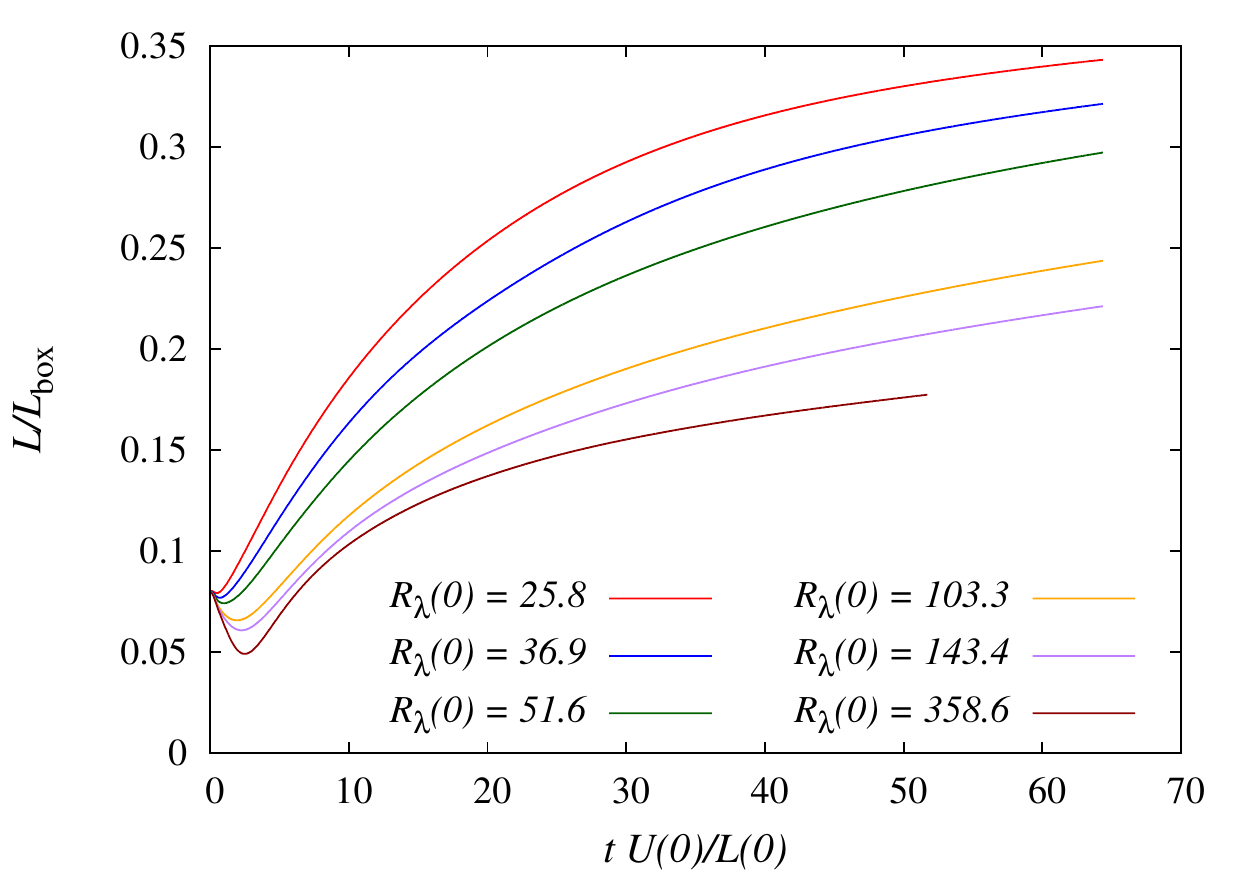}
 \end{center}
 \caption{The variation of the ratio of the integral scale to the box size 
 as time progresses.}
 \label{fig:decay_L}
\end{figure}

In view of the growing recognition of the importance of having the
integral length scale fully resolved, we show the time evolution of the
integral length scale during the decay in Fig. \ref{fig:decay_L}. The
integral length scale, $L$, increases during decay, as the energy becomes 
concentrated at larger and larger length scales. We wish to
check that the integral scale remains resolved during the simulation;
that is $L < L_\text{box} = 2\pi$. The figure shows the variation of
$L/L_\text{box}$ with time for a range of initial Reynolds numbers. The
ratio becomes smaller (i.e. better) as Reynolds number is increased.

\subsection{The occurrence of peak dissipation rate (at $t=t_\vep$) and transfer
rate (at $t=t_\Pi$)}

\begin{figure}[tbp]
 \begin{center}
  \includegraphics[width=0.6\textwidth]{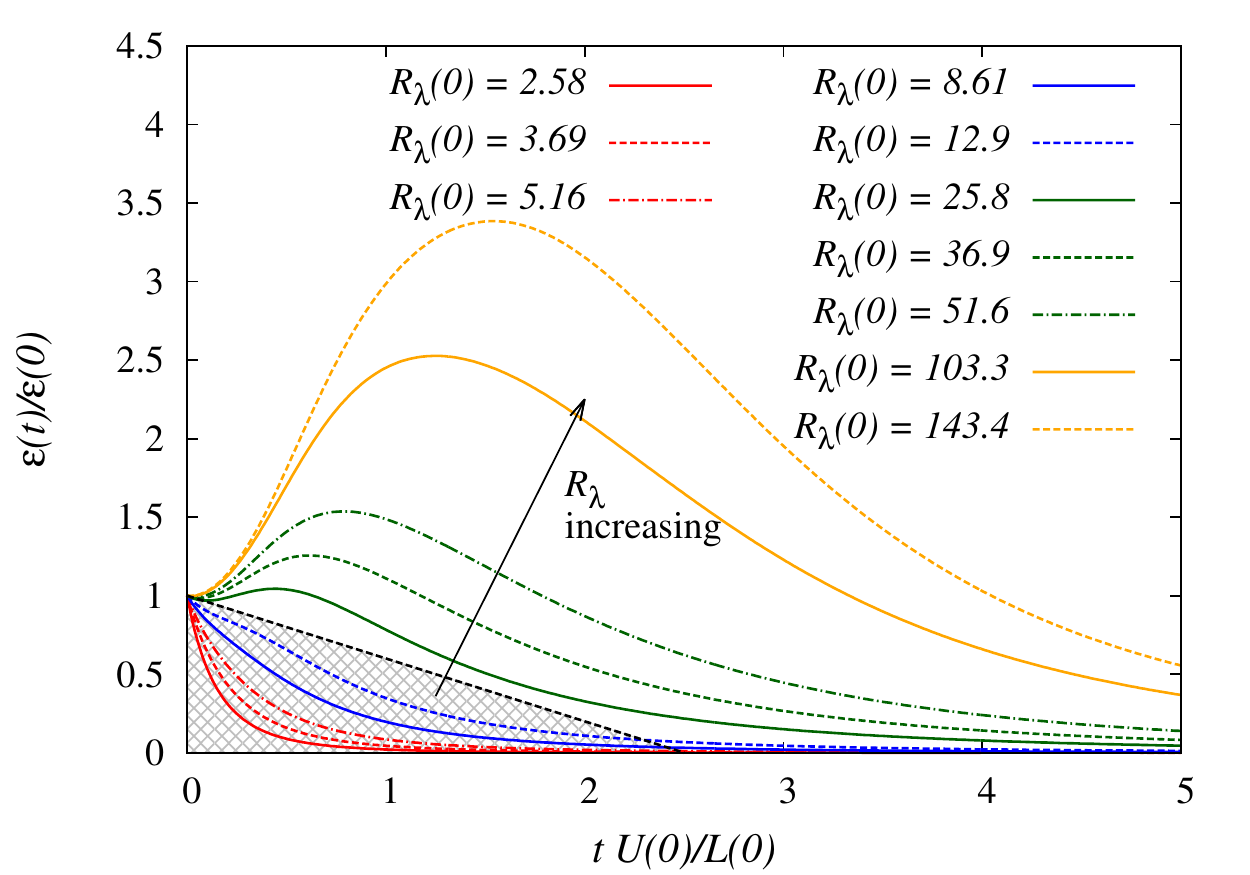}
 \end{center}
 \caption{Evolution of the dissipation rate $\varepsilon$ for a range 
 of simulations. For the lowest Reynolds numbers, as
 highlighted by the grey shading, there is no initial increase.}
 \label{fig:dissipation_decay}
\end{figure}

The existence of a  peak in the variation of any quantity with time
during the decay offers the possibility of a well-defined criterion,
which would allow the results of one investigation to be compared with
those of another. As seen in Fig. \ref{fig:dissipation_decay}, in those
of our simulations with $R_\lambda(0) > 25$, the dissipation rate
initially  increases and develops a peak. This peak value is easily
identified, so we denote the corresponding time to the peak by
$t=t_\vep$. As mentioned earlier, this time $t_\vep$ was the evolution criterion used for
decaying simulations by Fukayama \etal\ \cite{Fukayama00}.

\begin{figure}[tbp]
 \begin{center}
  \includegraphics[width=0.6\textwidth]{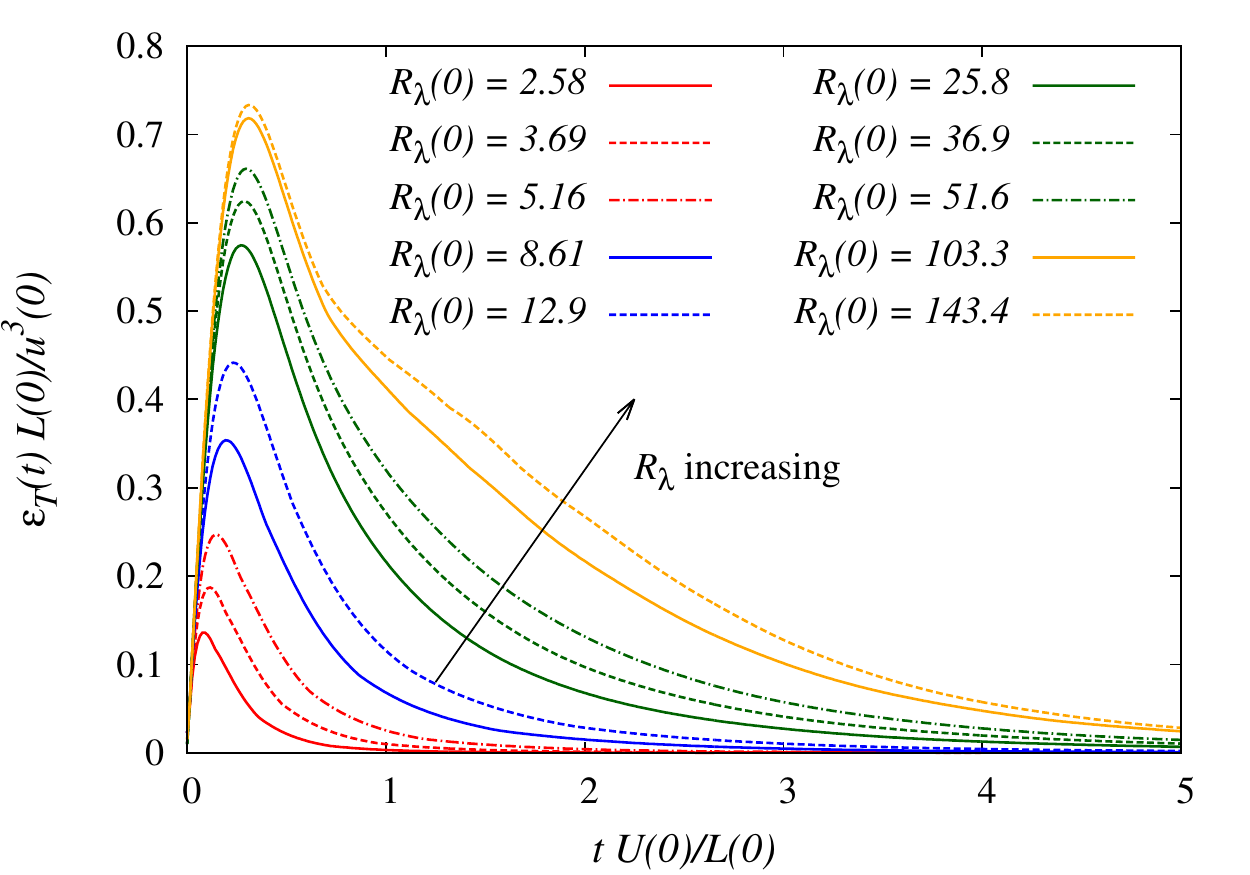}
 \end{center}
 \caption{Evolution of the maximum inertial flux $\varepsilon_T$ for a 
 range of simulations.}
 \label{fig:transfer_decay}
\end{figure}

However, for $R_\lambda(0) \lesssim 25$, dissipation dominates from the
start and there is no peak in $\varepsilon$. Instead, consider time
series for the maximum inertial flux, $\varepsilon_T$, as shown in
Figure \ref{fig:transfer_decay}. Here we see that all the curves have
peaks, and accordingly we may introduce a time $t=t_\Pi$ as the position
of the peak in the inertial flux.

\subsection{The occurrence of the peak skewness: at $t=t_S$}

\begin{figure}[tbp]
 \begin{center}
  \includegraphics[width=0.6\textwidth]{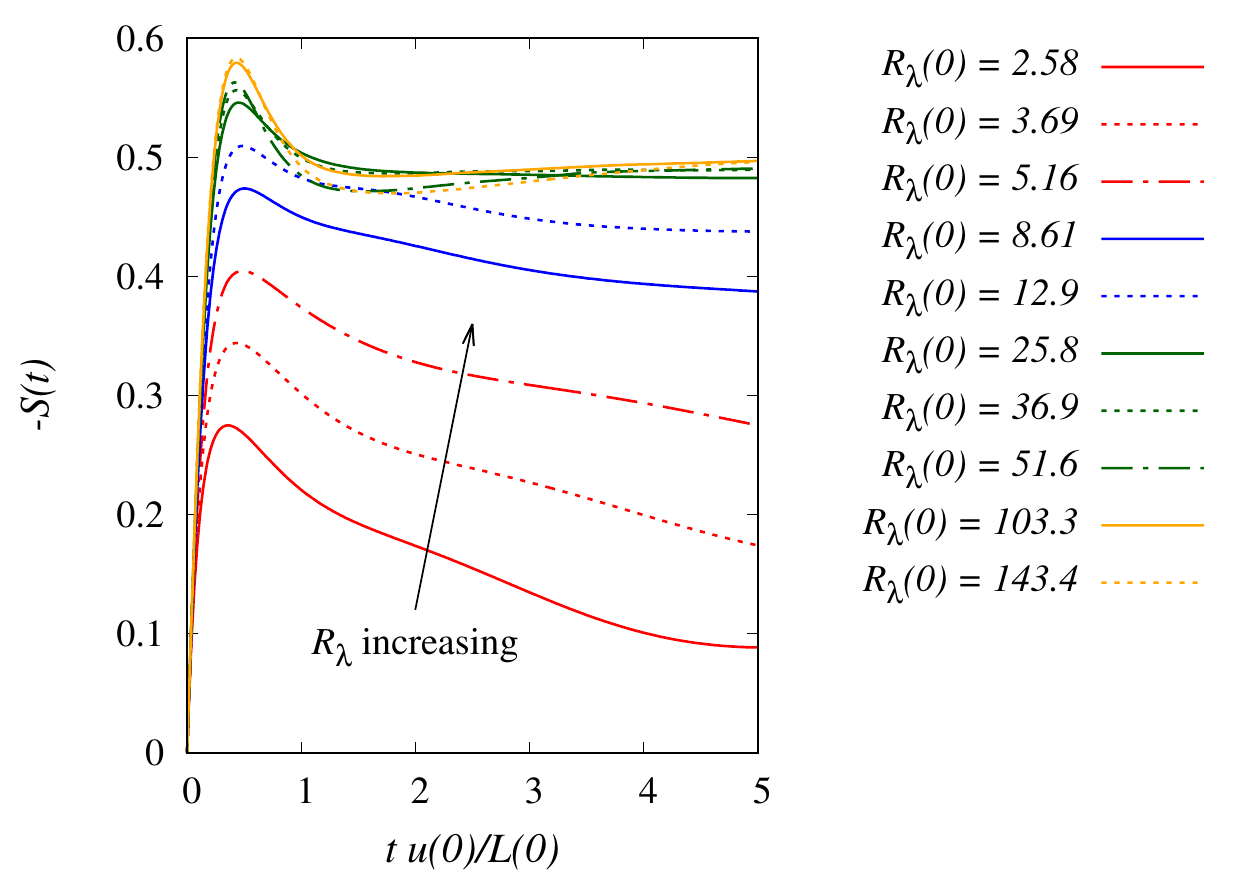}
 \end{center}
 \caption{Evolution of the velocity derivative skewness, $S(t)$. All 
 Reynolds numbers display a peak, with higher Reynolds numbers also later 
 developing plateaux for $t > 3\tau(0) = 3 L(0)/U(0)$.}
 \label{fig:skewness_decay}
\end{figure}

The velocity derivative skewness is often used as a criterion for fully
developed flow. Developed turbulence has a non-Gaussian probability
distribution. In a simulation the skewness starts from $S(0) = 0$, since
the  initial field is Gaussian, and grows with time as the nonlinear
interactions develop. We show our results for the skewness in Figure
\ref{fig:skewness_decay} as a function of time for various Reynolds
numbers. It may be seen that the skewness asymptotes towards a value of
$S\sim -0.5$ as both time and the Reynolds number increase. This is in
broad agreement with the results of other investigations as the skewness
has been found to be $S \sim -0.5$ for stationary turbulence: see
Ishihara \etal\ \cite{Ishihara09}; Machiels \cite{Machiels97a}; Vincent
and  Meneguzzi \cite{Vincent91}; Kerr \cite{Kerr85}; Gotoh \etal\
\cite{Gotoh02}; Jim\'enez \etal\ \cite{Jimenez93}; and Wang \etal\
\cite{Wang96}.

For large enough Reynolds numbers, a plateau develops  around $S \sim
-0.5$. The same set of Reynolds numbers which did not exhibit a  peak in
$\varepsilon$ are the ones which do not reach a plateau here. However,
for all $Re$, there is a peak, corresponding to the evolved time $t_S$.
Like $t_\Pi$, $t_S$ occurs very early in the decay for all Reynolds
numbers.

\subsection{Decay from an evolved field}

In an experiment to provide a more realistic starting point for
DNS of free decay, we used the data from a forced, stationary simulation
to set the initial conditions. That is, we started from an evolved field
which was taken from the $1024^3$ simulation with $R_\lambda = 335$. As
this data is, in effect, a solution to the \emph{forced} Navier-Stokes
equation, inevitably there is a transient after the forcing is `switched
off' as the system adapts to the lack of an energy input.

This was a straightforward matter, with the only problems arising from
the use of large data sets and the consequent long run times. We used an
ensemble of five initial fields (separated by one large eddy  turn-over
time, $\tau(0)$). It is of interest to note that $\tau(0) = L/U = 1.94s$
from the stationary simulation which is much longer than initial
turnover time for a Gaussian initial condition $\tau(0) = 0.777s$.

\begin{figure}[tbp]
 \begin{center}
   \includegraphics[width=0.6\textwidth,trim=5px 0 10px 0,clip]{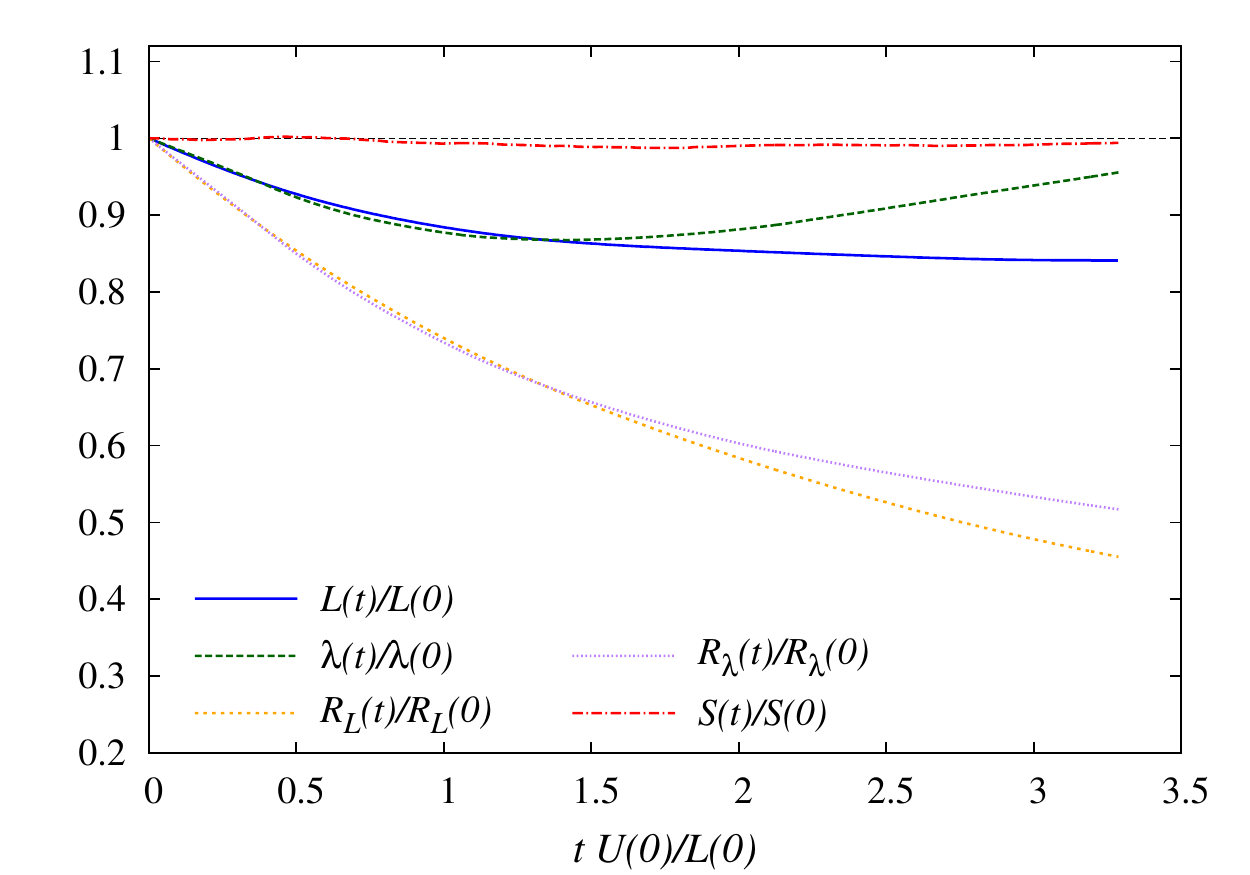}
 \end{center}
 \caption{Time evolution of length-scales, Reynolds numbers and skewness for 
 decaying turbulence with an evolved initial condition.}
 \label{fig:fdecay_params}
\end{figure}

The evolution of various parameters with dimensionless decay time is
shown in Figure \ref{fig:fdecay_params}. Evidently the normalised length
scales and Reynolds numbers decay from $t = 0$. The exception is the
skewness (which had already reached a stationary value of $S=0.55$)
which seems to adopt a slightly lower value, but does not vary
significantly.

\begin{figure}
 \begin{center}
  \includegraphics[width=0.6\textwidth,trim=5px 0 10px 
  0,clip]{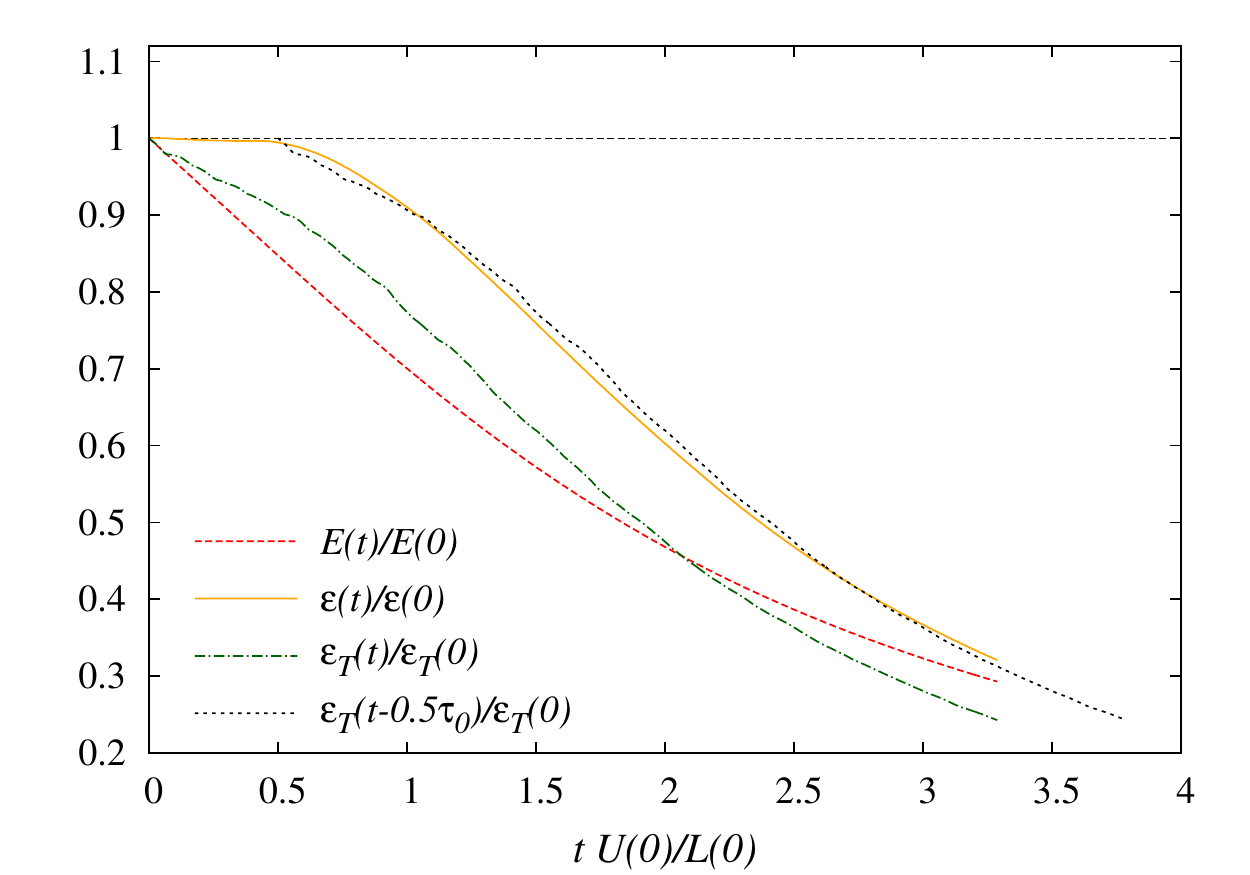}
 \end{center}
 \caption{Time variation of the total energy, dissipation rate and inertial 
 flux for decaying turbulence with an evolved initial condition. The dotted 
 line shows the transfer rate curve shifted to the right by $0.5\tau(0)$.}
 \label{fig:fdecay_rates}
\end{figure}

In Figure \ref{fig:fdecay_rates} we show the evolution of the
dissipation and transfer rates. We note that the total energy and
inertial flux, $\varepsilon_T$, start to decay  straight away. But there
is a period of about $\sim 0.5\tau(0)$ during which the dissipation rate
remains constant. This essentially measures the time it takes for energy
to pass  through the cascade, since during this time the dissipation
from high  wavenumbers is not `aware' of the change which has occurred
at low  wavenumbers.

We also note that the dissipation rate appears to mimic the transfer
rate. We tested this by shifting the $\varepsilon_T$ curve to the right
by  $0.5\tau(0)$, and we see $\varepsilon_T(t-0.5\tau(0)) \simeq 
\varepsilon(t)$. Or equivalently $\varepsilon(t+\Delta t) \simeq 
\varepsilon_T(t)$ with $\Delta t = 0.5\tau(0)$. This reinforces the
point above, that the time taken for the energy to pass through the
cascade is one half of the initial large eddy turnover time.

\section{A composite onset criterion based on both dissipation and energy transfer rates}

\begin{figure}[tbp]
 \begin{center}
  \includegraphics[width=0.6\textwidth]{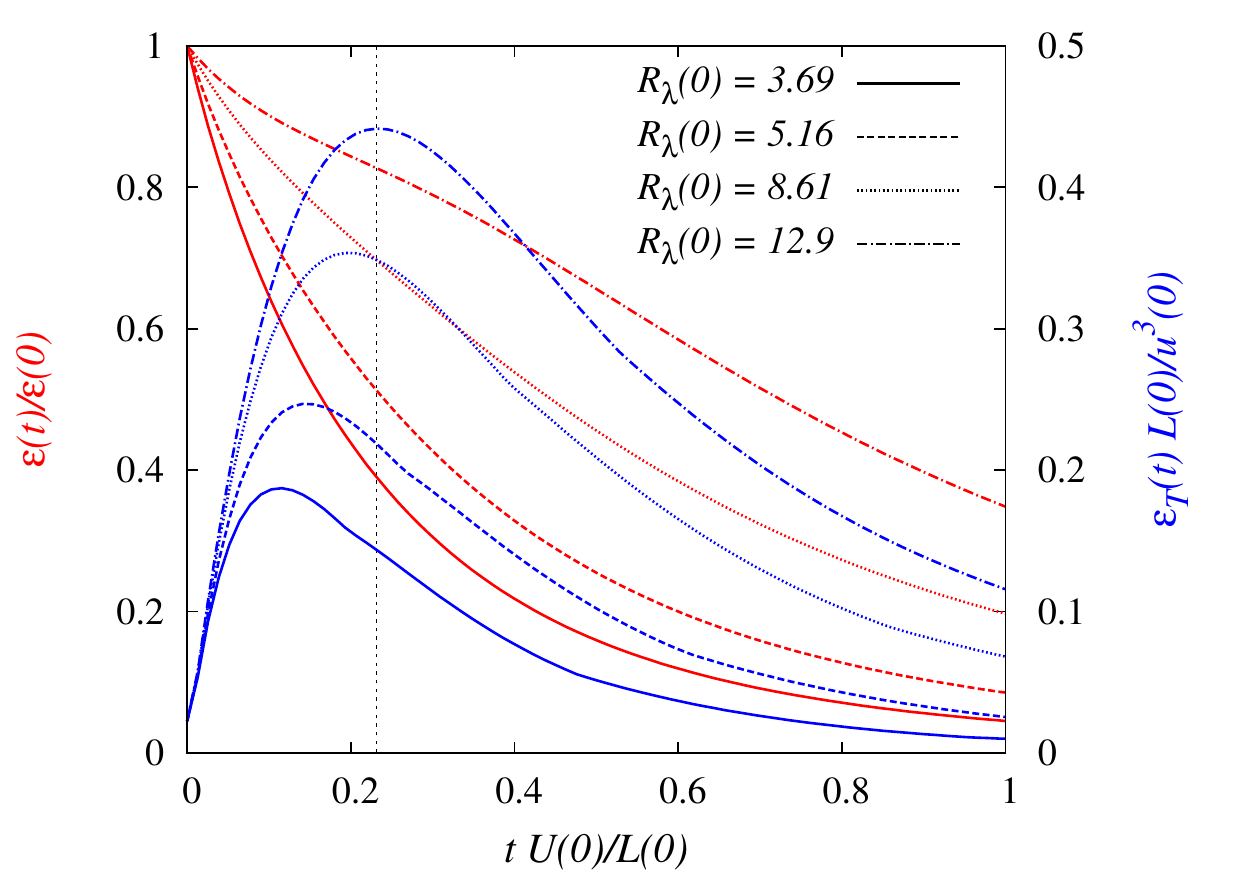}
 \end{center}
 \caption{Comparison of $\varepsilon(t)$ and $\varepsilon_T(t)$ for lower 
 Reynolds numbers. Peak in $\varepsilon_T$ can be seen to coincide with an 
 inflection in the dissipation rate. The vertical dotted line indicates the 
 position of $t_\Pi$ (corresponding to peak $\varepsilon_T$) for $\Rl=12.9$.}
 \label{fig:diss_transfer_decay}
\end{figure}

As we have seen, the curves for dissipation and energy transfer plotted 
against time display clear peaks, although in the case of the
dissipation rate, there is no peak for $\Rl \lesssim 25$. This leads us
to consider an alternative to using just either the dissipation or the
maximum inertial flux. In Figure \ref{fig:diss_transfer_decay}, we plot
both the dissipation and the maximum flux against time, for low initial
Reynolds numbers. In fact, the peak transfer coincides nicely with an
inflection point in  $\varepsilon$ for these low Reynolds number runs
which do not develop a peak in  the dissipation rate. The vertical
dotted line on the graph indicates the peak in $\varepsilon_T$ for
$\Rl=12.9$. This suggests the idea of a composite evolved time, which we
define as:
\begin{equation}
   \teP = \left\{ \begin{array}{ll}
    \tE & \text{if peak in }\varepsilon\text{ exists} \\
    t_\Pi         & \text{otherwise}
   \end{array} \right. \ .
  \end{equation}
That is, we use the time associated with the peak in the dissipation
rate curve, if it exists. Or, failing that, the peak in the curve of the
inertial flux.

\begin{figure}[tbp]
 \begin{center}
  \includegraphics[width=0.6\textwidth]{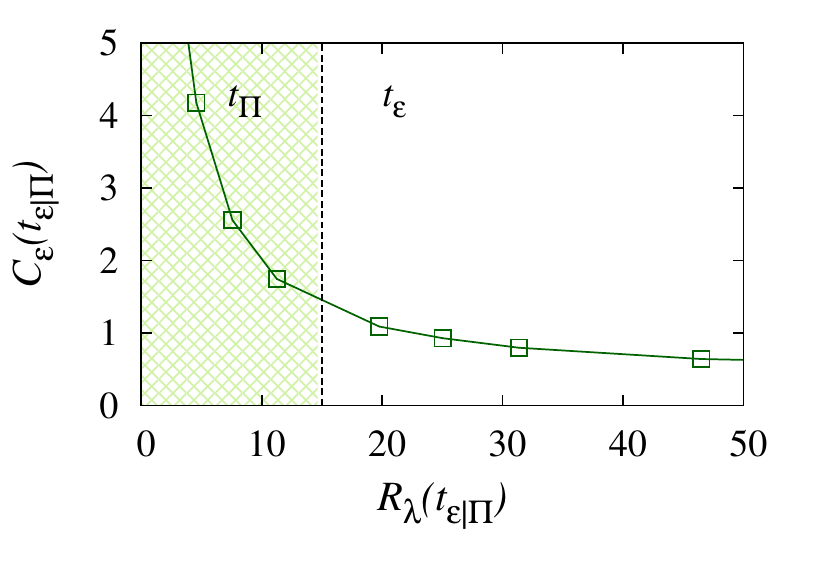}
 \end{center}
 \caption{Continuity of measured dimensionless dissipation rate between
two regimes}
 \label{fig:Continuity_Ceps}
\end{figure}

This provides us with a continuous evolved time. From Figure
\ref{fig:Continuity_Ceps} we see that there is no discontinuity as we go
from one regime to the other.

\section{Effect of choice of evolved time criterion on measurements of the
dimensionless dissipation}

As well as considering the traditional method of taking the onset of
power-law behaviour as the criterion for evolved time, we consider the
effects of the following criteria:
\begin{enumerate}
\item $t_{s}$ the time taken for the skewness to reach its peak value.
\item $t_{\Pi}$ the time taken for the inertial transfer rate to reach
its peak value.
\item $t_{\varepsilon}$ the time taken for the dissipation rate to reach
its peak value.
\item $t_{\varepsilon|\Pi}$ a composite time equal to
$t_{\varepsilon}$, if peak $\varepsilon$ exists; but equal to $t_\Pi$
otherwise. 
\end{enumerate}

\begin{figure}[tbp]
 \begin{center}
  \includegraphics[width=0.6\textwidth]{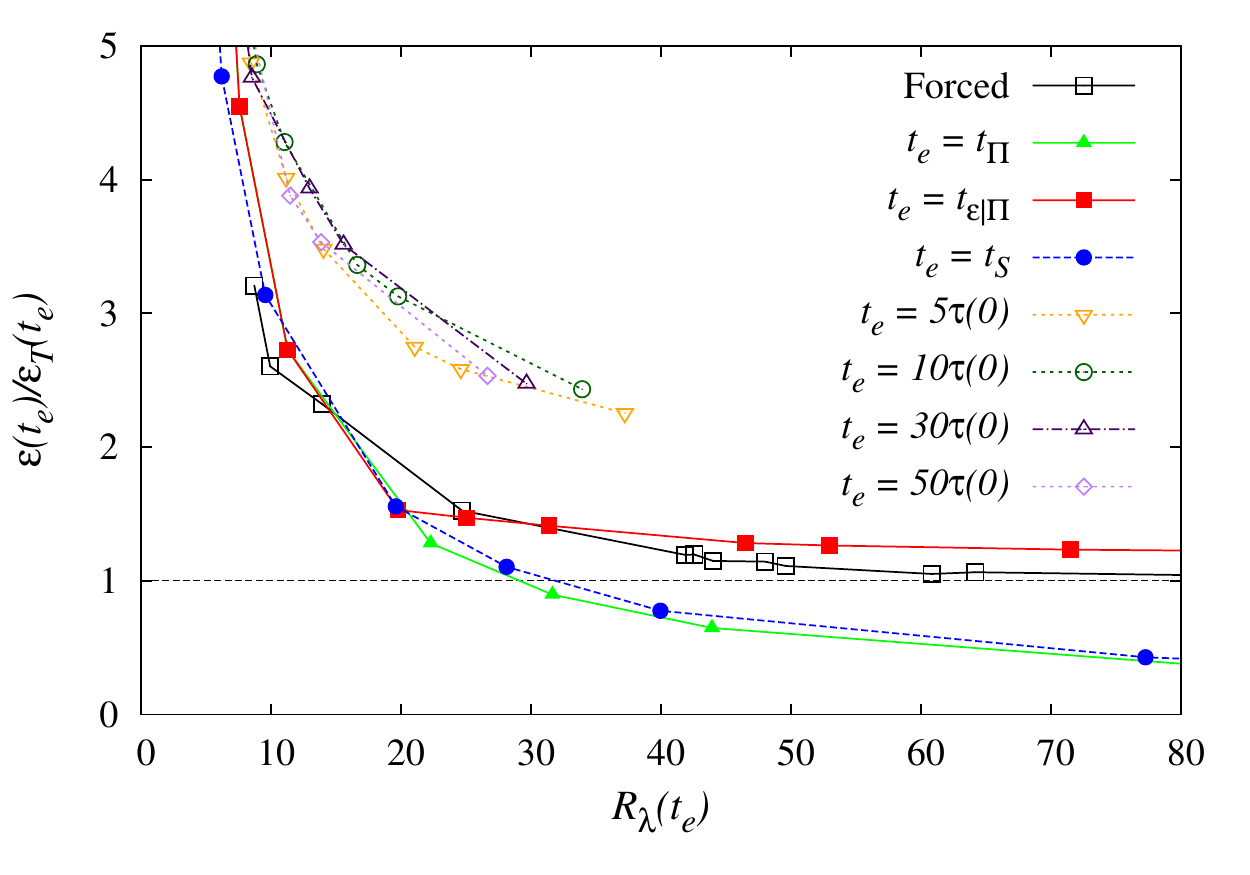}
 \end{center}
 \caption{The ratio $\varepsilon/\varepsilon_T$ measured using various 
 criteria for the evolved time, $t_e$. Results for forced, stationary turbulence shown for comparison.}
 \label{fig:decay_eps-pi}
\end{figure}

In Figure \ref{fig:decay_eps-pi} we plot the ratio
$\varepsilon/\varepsilon_T$ against Taylor-Reynolds number for  decaying
turbulence, using different criteria for the evolved time. We show
results taken at $t_e=\teP$, $t_e=t_\Pi$, and $t_e=t_S$, along with four
arbitrary times at various points during the decay, thus: $t_e =
5\tau(0)$, $t_e=10\tau(0)$, $t_e=30\tau(0)$, and $t_e=50\tau(0)$, where
$\tau(0)$ is the initial value of the eddy turnover time.  As the times
$\teP$ and $t_S$ occur much earlier in the decay than  power-law decay
of the total energy, it is interesting to to compare how the  choice of
evolved time affects measurements.

First let us consider what happens with forced turbulence. In this case,
energy enters at a rate $\varepsilon_W$. At the steady state there is a
balance of energy in and out: $\varepsilon = \varepsilon_W$. At low
$Re$, some energy is dissipated by wavenumbers below the inertial range,
so we expect to find $\varepsilon_T < \varepsilon$. As the Reynolds
number is increased, we expect that $\varepsilon_T \to \varepsilon$.
 
The situation is different for decaying turbulence but also depends on
the Reynolds number. Let us begin with the case of high $Re$. We can
measure $\varepsilon_T(t) > \varepsilon(t)$ in the initial  transition
period. Due to the finite transit time for energy transfer through the cascade, energy 
transferred at time $t$ will be dissipated at time $t+\Delta t$. Or 
$\varepsilon(t+\Delta t) = \varepsilon_T(t)$.  The turbulence is
decaying, so $\varepsilon(t + \Delta t) < \varepsilon(t)$, thus we must
have $\varepsilon_T(t) < \varepsilon(t)$. Hence the time $\tE$ is the
border between the cases $\varepsilon_T(t) > \varepsilon(t)$ and
$\varepsilon_T(t) < \varepsilon(t)$. One would therefore expect to measure
$\varepsilon_T(\tE) =  \varepsilon(\tE)$.
 
Turning to the  case of decaying turbulence at low $Re$: we would still
measure $\varepsilon_T(t) > \varepsilon(t)$ in the  initial transition
period. For $t \geq \tE$, we now must have $\varepsilon(t) > \varepsilon_T(t)$. 
The peak dissipation rate is no longer associated with
equality of  transfer and dissipation, but this is due to finite $Re$.

We may summarise the results of  Fig. \ref{fig:decay_eps-pi} as follows:  

\begin{enumerate}

\item Skewness: $\varepsilon_T(t_S) > \varepsilon(t_S)$ indicates we are in 
transition period. 

\item Combined: Measure $\varepsilon_T(\teP) < \varepsilon(\teP)$ for all 
Reynolds numbers, but could be asymptoting to unity.

\item Power-law decays: All in agreement with one another; do not asymptote 
to unity.

\item $\lambda \propto t^{1/2}$: Measurements at $t_e = 30\tau(0),\ 
50\tau(0)$.

\end{enumerate}

\begin{figure}
 \begin{center}
  \includegraphics[width=0.6\textwidth]{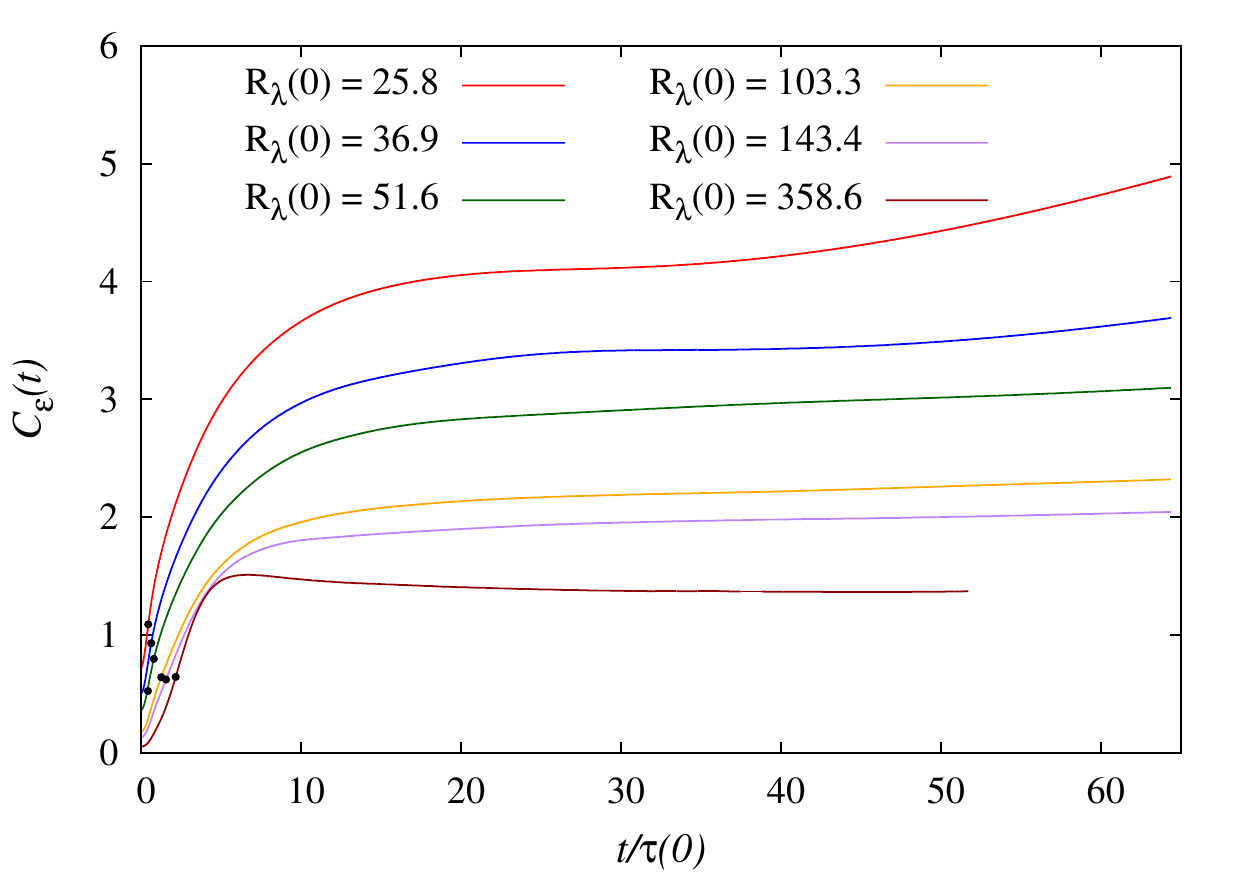}
 \end{center}
 \caption{Time evolution of the dimensionless dissipation rate, $\Ceps(t) = 
 \varepsilon(t) L(t)/U^3(t)$, plotted against decay time scaled on the
initial eddy turnover time $\tau(0)$. Note that the solid black circles
are the values of $t_{\varepsilon|\Pi}$.}
 \label{fig:decay_Ceps_TS}
\end{figure}

We may look more deeply into the measurement of the
dimensionless dissipation coefficient by looking at its dependence on
time. This illustrates some of the problems involved. We refer to Fig.
\ref{fig:decay_Ceps_TS}, which shows the time evolution of $\Ceps(t)$
for various values of the Taylor-Reynolds number. The measurement time
$\teP$, as indicated by solid points for the various $\Rl$, occurs very
early in the decay while $\Ceps$ is strongly time-dependent.

We note that $\Ceps(t)$ develops a plateau from around $t \simeq 10\tau(0)$.
But, while $\Ceps(t)$ then remains constant, the Reynolds number still decays. 
Therefore, the same value of $\Ceps(t)$ corresponds to different 
$R_\lambda(t)$ and a plot of $\Ceps(t)$ against $R_\lambda(t)$ for different 
$t$ will have their curves shifted.

The behaviour $\Ceps(t) \sim \text{constant}$ observed in Fig.
\ref{fig:decay_Ceps_TS} for a period of the decay requires:

\begin{equation}
   \frac{L}{\lambda} \sim R_\lambda \ , \qquad 
   \left(\frac{L}{\lambda}\right)^2 \sim R_L \ , \quad \text{or}\qquad R_L 
   \sim R_\lambda^2 \ .
\end{equation}

\begin{figure}
 \begin{center}
  \includegraphics[width=0.6\textwidth]{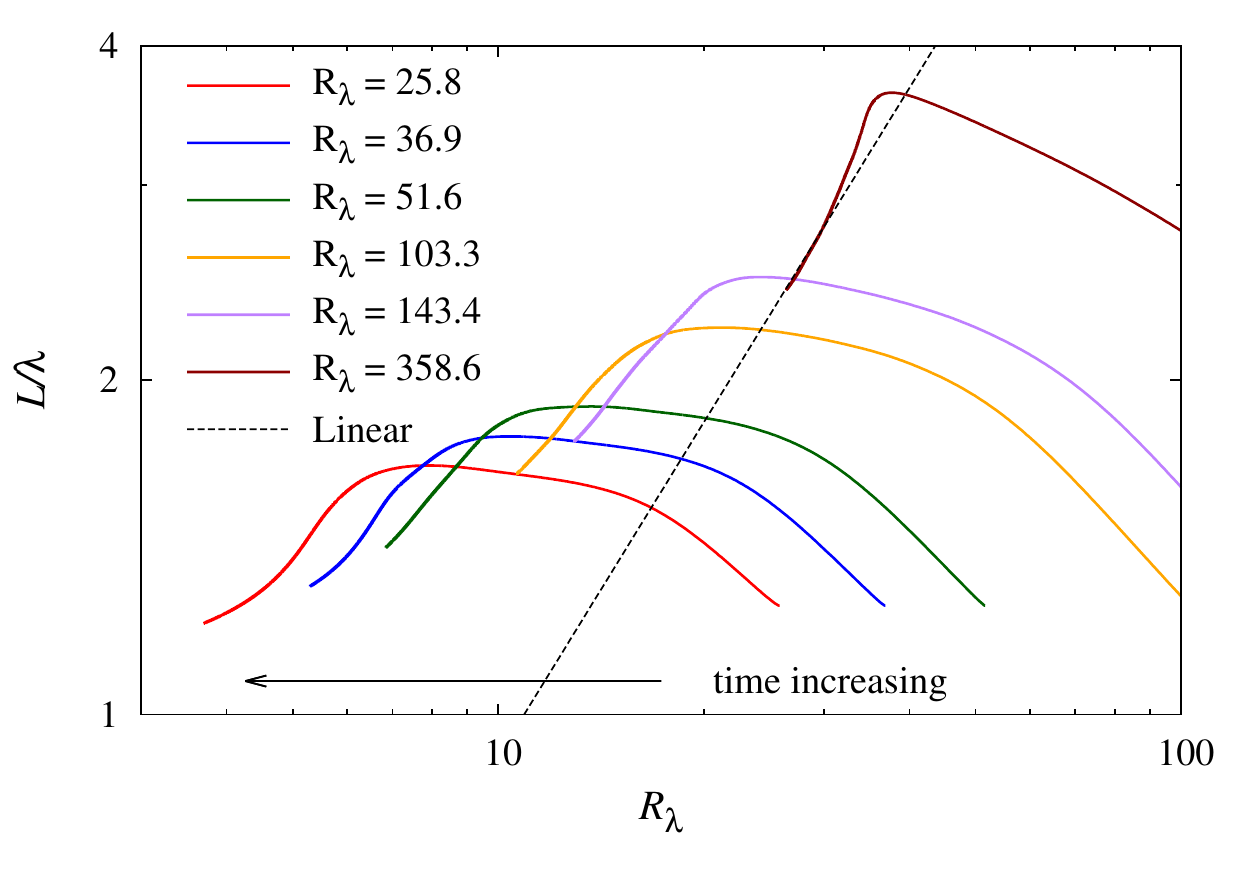}
 \end{center}
 \caption{Variation of the ratio of integral scale $L$ to microscale 
 $\lambda$ during the decay, as a function of Taylor-Reynolds number.
 Since $R_\lambda$ falls off during the decay, time runs from right to left.  
 Linear behaviour for each curve is shown by a region parallel to the dotted 
 line.}
 \label{fig:decay_Ll_Rl}
\end{figure}

Fig. \ref{fig:decay_Ll_Rl} shows $L/\lambda$ plotted against
$R_\lambda$ during the decay. We may observe a region at lower Reynolds
number $\Rl$ where the relationship is approximately linear (as
indicated by the dashed line). This behaviour was found experimentally
for regular grids by Valente and Vassilicos \cite{Valente12b} for low
$R_\lambda$ far downstream, and followed a region where $L/\lambda \sim
\text{constant}$, in which case $\Ceps \sim 1/R_L \sim 1/R_\lambda$.
Their results for a fractal grid show good agreement, but far fewer points
were plotted (see the open squares in their figure 5).

\begin{figure}
 \begin{center}
  \includegraphics[width=0.6\textwidth]{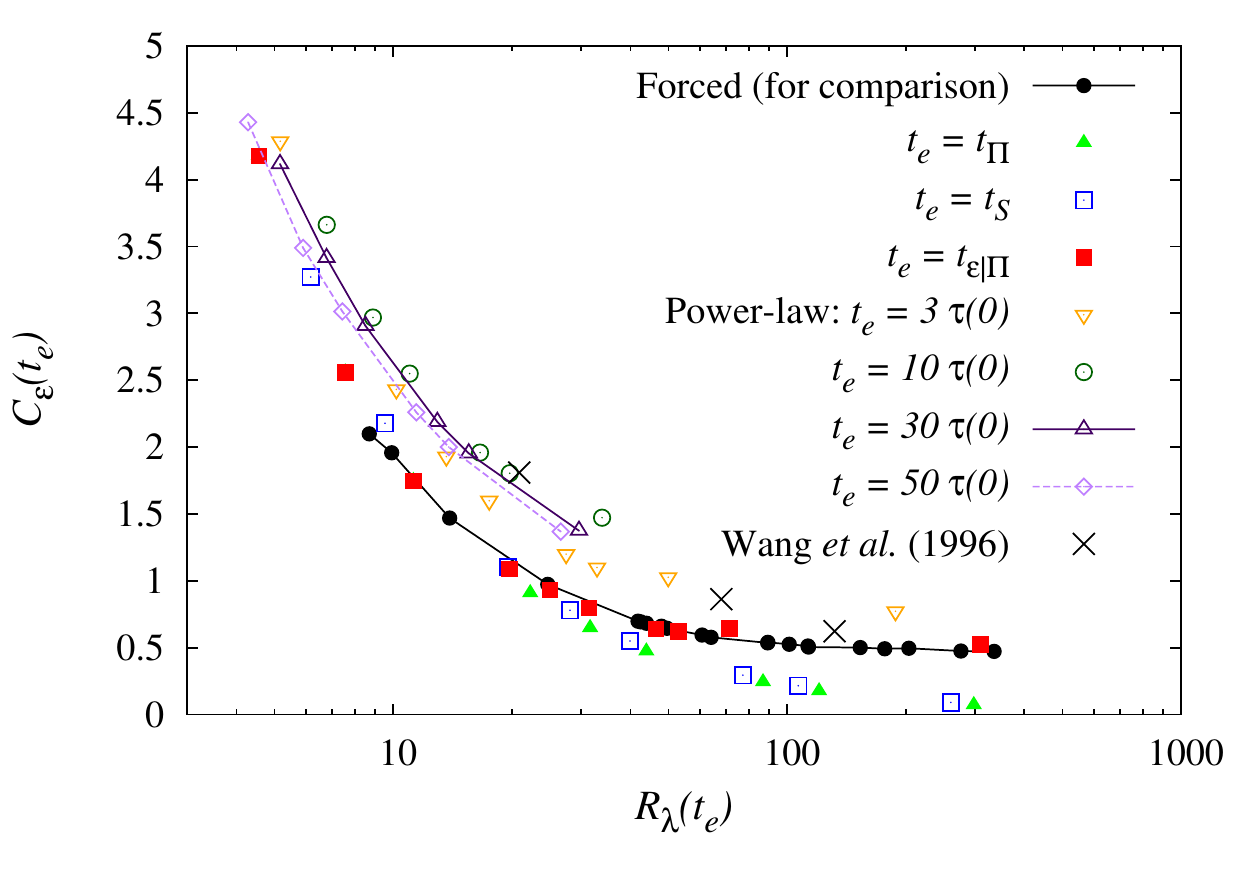}
 \end{center}
 \caption{The Reynolds number variation of the dimensionless dissipation 
 rate for decaying turbulence. Plotted for various choices of the evolved 
 time $t_e$, along with results for the forced case and decay data from Wang \etal\ 
 \cite{Wang96}, for comparison.}
 \label{fig:decay_Ceps}
\end{figure}

Lastly, we consider the question of how the usual curves of $\Ceps$
against $\Rl$ are affected by our choice of onset criterion. This is shown
in Fig. \ref{fig:decay_Ceps}, which makes a comparison of $\Ceps(t_e)$ against
$R_\lambda(t_e)$ for different evolved time criteria. When we use the peak
skewness, $t_S$ or the peak inertial transfer time $t_\Pi$, the
dimensionless dissipation rate appears to approach zero as $R_\lambda$
increases. However, if we take our preferred criterion, based on the dissipation,
$\teP$ then the dimensionless dissipation rate appears
to match the forced case (which is plotted for comparison).
  
If we go to later measurement times, then we find very different
behaviour. For choices of $t_e$ in the range $3\tau(0)\leq t_e \leq 50
\tau(0)$, we find that the curves cluster together and follow a similar
profile to the forced case, only translated up the $\Ceps$-axis.
   
In all, we conclude that the asymptote for decaying turbulence is in the range $0 \leq
\Cinf(t_e) \lesssim 1.2$, depending on the choice of evolved time.

\section{Conclusion}

We have been mainly concerned with the way in which choice of an
evolution time $t_e$ affects the shape of the curve of $\Ceps$ when
plotted against $\Rl$. Figure \ref{fig:decay_Ceps} is the key result
here. It is clear that different choices of evolution time can produce
radically different results. A failure to recognise this presents a
problem for the comparison of one investigation with another. Indeed,
this degree of variation is quite sufficient to explain the spread of
results commonly found in surveys.

We would suggest that meaningful comparisons would need one to compare
the detailed evolution (with time or space) of the dissipation rate (see
Figure \ref{fig:dissipation_decay}) or the dimensionless dissipation
rate (see  Figure \ref{fig:decay_Ceps_TS}). It would be more
satisfactory if a consensus could arise regarding a choice of onset
criterion, although we cannot dismiss the possibility that separate
criteria might be needed for different statistical quantities. 

Our preferred criterion $t_{\varepsilon|\Pi}$ has some physical appeal
in that it is determined from primary statistical quantities rather
than the secondary characteristic of an apparent power law decay.
Moreover, although it is a composite form, it displays continuous
behaviour as a criterion in a plot of dimensionless dissipation versus
Taylor-Reynolds number as seen in Fig. \ref{fig:Continuity_Ceps}. Thus
it links the low-Reynolds number flows with the high-Reynolds number
flows, as defined in Fig. \ref{fig:dissipation_decay}. It should perhaps
be emphasised that these low-Reynolds number flows cannot be dismissed as
being nonturbulent, as we have previously reported for low Reynolds numbers
in \emph{forced} turbulence \cite{McComb15b}. From the results shown in Figures
\ref{fig:transfer_decay} and \ref{fig:skewness_decay}, it can be seen
that even at these low Reynolds numbers, there is a development away
from the initial Gaussian state, with the growth of inertial transfer
and skewness from zero, to reach a peak value.

However, if we are to make meaningful comparisons between decaying and
stationary turbulence, then a more fundamental justification is needed.
As mentioned in Section 1, recently McComb and Fairhurst
\cite{McComb18a} showed by means of asymptotic expansion of structure
functions in powers of the inverse Reynolds number, that the time
derivative in the Karman-Howarth equation has an irreducible remainder
which does not vanish in the limit of large Reynolds numbers or at
particular length scales. Accordingly, they concluded that Kolmogorov's
hypothesis of local stationarity cannot be correct. The question then
arises: how large is the error if we neglect the term? In principle,
this can be answered by a direct comparison of $\Cinf$ in both forced
and decaying flows. But for this comparison to be meaningful we need an
agreed definition (and hence value) of $t_e$.

If we insist on our composite criterion for the purposes of studying the
dimensionless dissipation, then we have to accept that there is a second
transition (at a later time) to power-law behaviour. This in itself
should not be too surprising. For instance, some simulations of forced
turbulence have shown a transition to turbulence from a Beltrami flow at
a specific Reynolds number, followed by a second transition to scaling
behaviour at a higher Reynolds number \cite{McComb15b}. Of course our
presentation here is purely in terms of physical reasoning. Ideally the
problem could be solved mathematically as an example of a phase
transition. However, even although the study of hydrodynamic stability
has been going on for 140 years, and is now an immense field of activity
(see the superb review by Zhou \cite{Zhou17}), we are not aware of any
possibility of a mathematical theory of the development of isotropic
turbulence in free decay.

We conclude by restating our general position. We are studying the decay
of the (initially) random motion of an incompressible, viscous fluid. We
extend our analysis by using DNS, and our method of doing this is part
of a long established general paradigm. Thus, when we explore the possibility
of an alternative starting point, it is more from the point of view of
statistical physics than from any conviction that there might be
something unsatisfactory about our initial spectra.  Because in this
paradigm, the qualitative (and quantitative) performance of the
simulations, in terms of statistical parameters, spectra, and fluxes, is
very well established. Morever, the other investigations that are
summarised in our Figure 1 are also part of the same paradigm.
Accordingly, all our results are applicable to these investigations, for
we are comparing like with like.

\section*{Acknowledgement}
The authors thank Professor Arjun Berera for many stimulating
discussions of this work. SRY is grateful for the support of an STFC
studentship while this work was undertaken, and is currently supported by
the UK EPSRC grant EP/N028694. Simulations were performed on the Eddie HPC
cluster hosted by the Edinburgh Parallel Computing Centre (EPCC).
Data are available online
[http://dx.doi.org/10.15129/64a4a042-7d0d-48ce-8afa-21f9883d1e84].

\newpage

\appendix

\section{Results for power-law behaviour from measurements of the local slope}

This is a confused and rather controversial topic.  For recent reviews
of the subject, we suggest the books \cite{Davidson04} and
\cite{McComb14a}. Here, we give a brief introduction to the topic
and present a summary of representative values of decay exponents in
Table \ref{tbl:exponent_summary}. In this way, it can be seen how our
own results fit in with the rest of the field. 

The topic is still dominated by the classic theories of Kolmogorov
\cite{Kolmogorov41c}, with $n=10/7 \sim 1.43$ and Saffman
\cite{Saffman67a}, with $n=6/5 \sim 1.2$. These are rival theories,
either of which may apply, depending on the initial conditions
determining whether the Loitiansky integral or the Saffman-Birkhoff
integral is an invariant. The former case is sometimes referred to as
\emph{Batchelor turbulence} and the latter as
\emph{Saffman turbulence}.\footnote{Indeed Batchelor \cite{Batchelor48}
also predicted $n=5/2$, but that was
for the final period of the decay, when viscous forces become dominant.}
This situation can be understood in terms of the infrared behaviour of
the energy spectrum $E(k,t)$. As is well known, this can be written as a
Taylor polynomial at small wavenumbers, thus:
\[
E(k,t)= E_2(t)k^2 + E_4(t)k^4 + \dots ,
\]
where $E_2(t)$ is the Saffman-Birkhoff integral and $E_4(t)$ is the
Loitsiansky integral. In fact a recent study of this problem has shown
that $E_2(t)=0$ is an exact result \cite{McComb16a}, and hence the
Saffman-Birkhoff integral is zero.

The most significant (relatively) modern experimental studies of this
topic were probably those due to Comte-Bellot and Corrsin in 1966 and
1971 \cite{Comte-Bellot66} and \cite{Comte-Bellot71}. These established
values of the decay exponent in the range $1.16 \leq n \leq 1.37$. This
was followed by Mohamed and LaRue \cite{Mohamed90} and later by Skrbek
and Stalp \cite{Skrbek00}. The first of these investigated the effect of
initial conditions, and concluded that they only affected the decay
coefficient and not the exponent or the virtual origin. A particularly
interesting feature of the work by Skrbek and Stalp was their
investigation of the relationship between the finite size of the
test-section and the largest eddies. They discussed a \emph{saturation
effect} in which the nature of the decay changed after the largest
eddies had grown to the same size as the `box'. This is now being
investigated in numerical simulations too, where finite size and
finite Reynolds number effects are being studied: see the recent work by 
Thornber \cite{Thornber16}, and Meldi and Sagaut \cite{Meldi17}. 

\begin{figure}
 \begin{center}
  \includegraphics[width=0.6\textwidth]{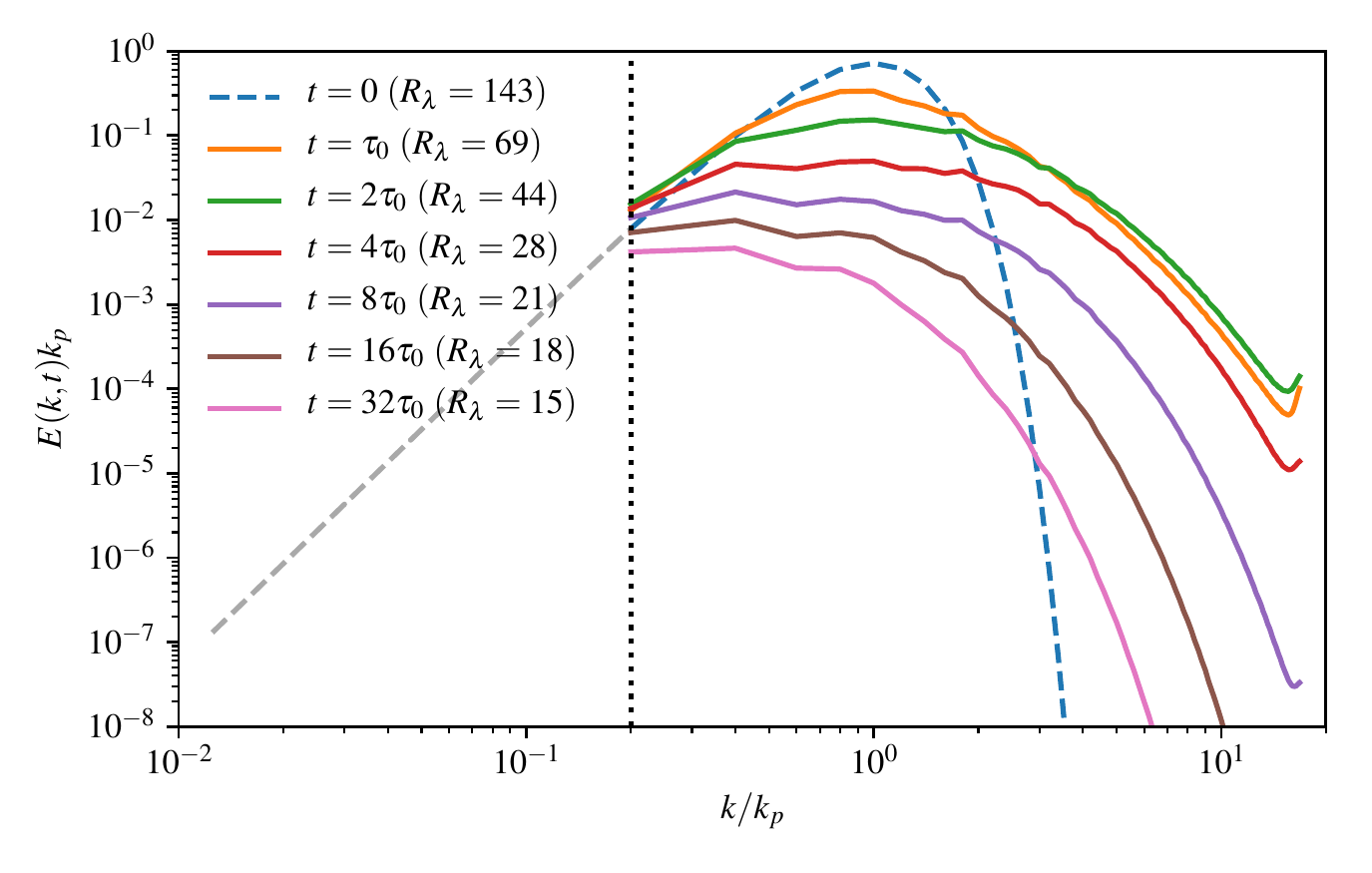}\\
  \includegraphics[width=0.6\textwidth]{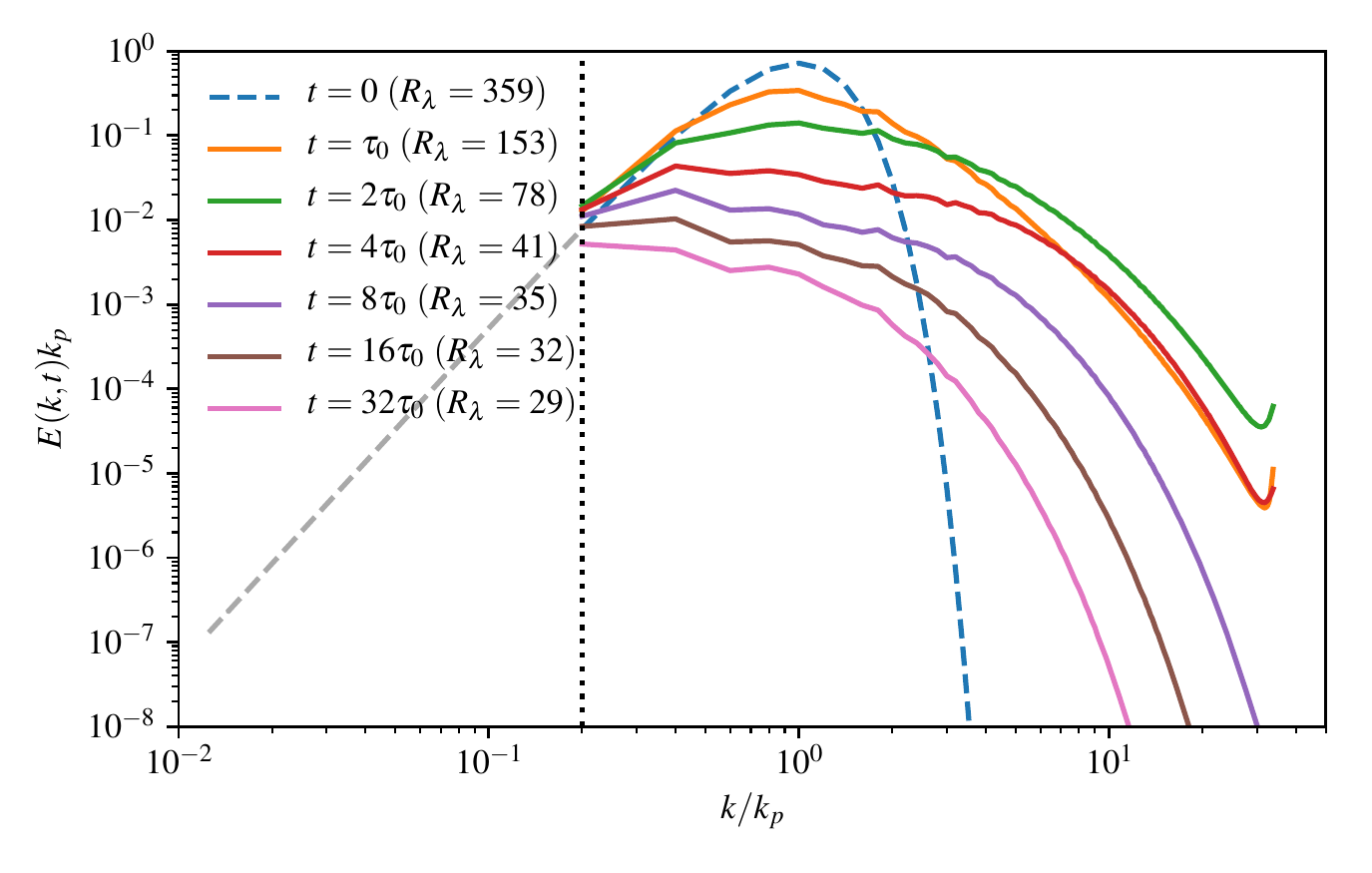}
 \end{center}
 \caption{Time evolution of the energy spectrum for one of our $256^3$
 simulations (top) and our $512^3$ simulation (bottom),
 plotted for comparison to Figure 4 from Ishida \textit{et al.}
 \cite{Ishida06}.}
 \label{fig:spectra}
\end{figure}

We may also mention the study of the effects of increasing the
wavenumber corresponding to the peak value of the initial spectrum by
Ishida, Davidson and Kaneda \cite{Ishida06}. This investigation uses an
initial spectrum similar to our equation \eqref{eq:intial_spec}. Custom
and practice normally dictate that this wavenumber (denoted by $k_p$ in Ishida
\emph{et al.}) is taken to be as small as possible in order to allow
turbulence to develop which is characteristic of the Navier-Stokes
equations, rather than of the initial conditions. It is not simply a
increase in the resolution of the integral lengthscale. However, Ishida
\emph{et al.} consider values of $k_p$ up to 80, as compared with our
equivalent of $k_p=5$, in order to study the low-$k$ behaviour of the
spectrum. Here we replot our results for the time evolution of the energy
spectrum as Figure \ref{fig:spectra} in order to make a comparison with
Figures 4, 7 and 9 from Ishida \emph{et al}. It may be seen that the
behaviour is quite similar. In addition, our results for the energy decay
exponent in Figure \ref{fig:decay_exp_lambda} may be seen to be in agreement
with those of Ishida \emph{et al.} as given in their Figures 3 and 6.
However, Ishida \emph{et al.} do not present any results for
transfer spectra or skewness factor to indicate that their turbulence is
well-developed, nor any data for the dissipation rate, velocity, or
lengthscales in order to facilitate a comparison of values of
$C_\varepsilon$.

More recent experimental studies include those of Vassilicos and
co-workers
\cite{Seoud07}-\nocite{Mazellier08}\nocite{Valente11a}\cite{Valente12b}
and by Krogstad and Davidson \cite{Krogstad10}. The latter measured
decay exponents in a particularly large wind tunnel and concluded that,
when they made allowance for a weak decay of the dimensionless
dissipation rate, the turbulence was
of the Saffman type. They also investigated turbulence generated by
multiscale grids \cite{Krogstad11}, and concluded that the resulting
decay was almost identical to that behind conventional grids. This was
in contrast to what had been found by Vassilicos and co-workers. A
subsequent paper by Valente and Vassilicos \cite{Valente11a} reaffirmed
the untypical properties of turbulence generated by multiscale grids.
Decay exponents for all three investigations are listed in Table
\ref{tbl:exponent_summary}.

\begin{table}
\begin{center}
\begin{tabularx}{\textwidth}{llX}
 Reference & $n$ & Comments \\
 \hline \hline
 Kolmogorov \cite{Kolmogorov41c} & $n = 10/7$ & Notes that 
 \cite{VonKarman38} found $E \propto t^{-n}$ but did not find $n$ \\
 Batchelor \cite{Batchelor48} & $n = 5/2$ & In the final period of the decay 
 \\
 Saffman \cite{Saffman67a} & $n = 6/5$ \\
 \hline
 Mohamed and & $n = 1$ and $1.16 \leq n \leq 1.37$& Review of experimental 
 results\\
 LaRue \cite{Mohamed90} & $n = 1.42,\ 1.33,\ 1.10,\ 0.95$ & Virtual origin 
 $x_0/M_u = 0,\ 2,\ 4,\ 6$ \\
  & $n = 1.24 \leq n \leq 1.33$ & Optimised origin $x_0$ --- table 4.\\
  \hline
    Yu \etal\ \cite{Yu05}  & $n = 1.38$--1.85 & LB: $R_\lambda =  
    2.3$--22.5 $(E(k)\sim k^4)$ \\
    \cite{Mansour94} & $n = 1.1$--1.52 & DNS: $R_\lambda =$ 0--30
    $(E(k)\sim k^2)$ 
     \\
    \cite{Huang94} & $n = 1.0$--3.0 & DNS: $R_\lambda = 10$--50
    $(E(k)\sim k^2)$ 
    \\
    \cite{Mohamed90} & $n = 1.285$--1.309 & Exp: $R_\lambda = 28.4$--43.9
    $(E(k)\sim k^2)$ \\
    \cite{Smith93} & $n = 1.3$--1.8 & Exp: $R_\lambda = 4.4$--5.4
    $(E(k)\sim k^2)$ 
    \\
    \hline
    Krogstad and & $1.15 < n < 1.29$ & Comte-Bellot and Corrsin (1966) \\
    Davidson \cite{Krogstad10} & $n \sim 1.34$ & Warhaft and Lumley (1978) 
    \\
    & $1.09 < n < 1.19$ & Find an average of $n = 1.13 \pm 0.02$\\
    \hline
    Valente and & $n = 1.25$--1.36 & Regular grid \\
    Vassilicos \cite{Valente11a} & $n = 1.93$--2.57 & Fractal grid \\
    \hline
    Krogstad and
    & $n = 1.13 \pm 0.02$ & Conventional grid (regression) \\
     Davidson \cite{Krogstad11} & $n = 1.14 \pm 0.02$ & Conventional grid 
     (local slope) \\
    & $n = 1.17 \pm 0.04$ & Conventional grid (maximum decay) \\
    & $n = 1.12 \pm 0.02$ & Fractal grid 1 (regression) \\
    & $n = 1.17 \pm 0.02$ & Fractal grid 1 (local slope) \\
    & $n = 1.19 \pm 0.03$ & Fractal grid 1 (maximum decay) \\
    & $n = 1.25 \pm 0.02$ & Fractal grid 2 (regression) \\
    & $n = 1.25 \pm 0.02$ & Fractal grid 2 (local slope) \\
    & $n = 1.23 \pm 0.03$ & Fractal grid 2 (maximum decay) \\
  \hline
  Our DNS data & $n = 1.49$ & $R_\lambda(0) = 143.4$ \\
   & $n = 1.35$ & $R_\lambda(0) = 358.6$
\end{tabularx}
 \end{center}
 \caption{A summary of values, both theoretical and experimental, of the 
 decay exponent $n$ such that $U^2 \propto t^{-n}$, from the literature. 
 Note that `LB' stands for Lattice
Boltzmann, `DNS' stands for Direct Numerical Simulation, and `Exp'
stands for experimental.}
 \label{tbl:exponent_summary}
\end{table}

\subsection{Onset criteria based on power-law decay}

\begin{figure}[tbp]
 \begin{center}
  \includegraphics[width=0.6\textwidth]{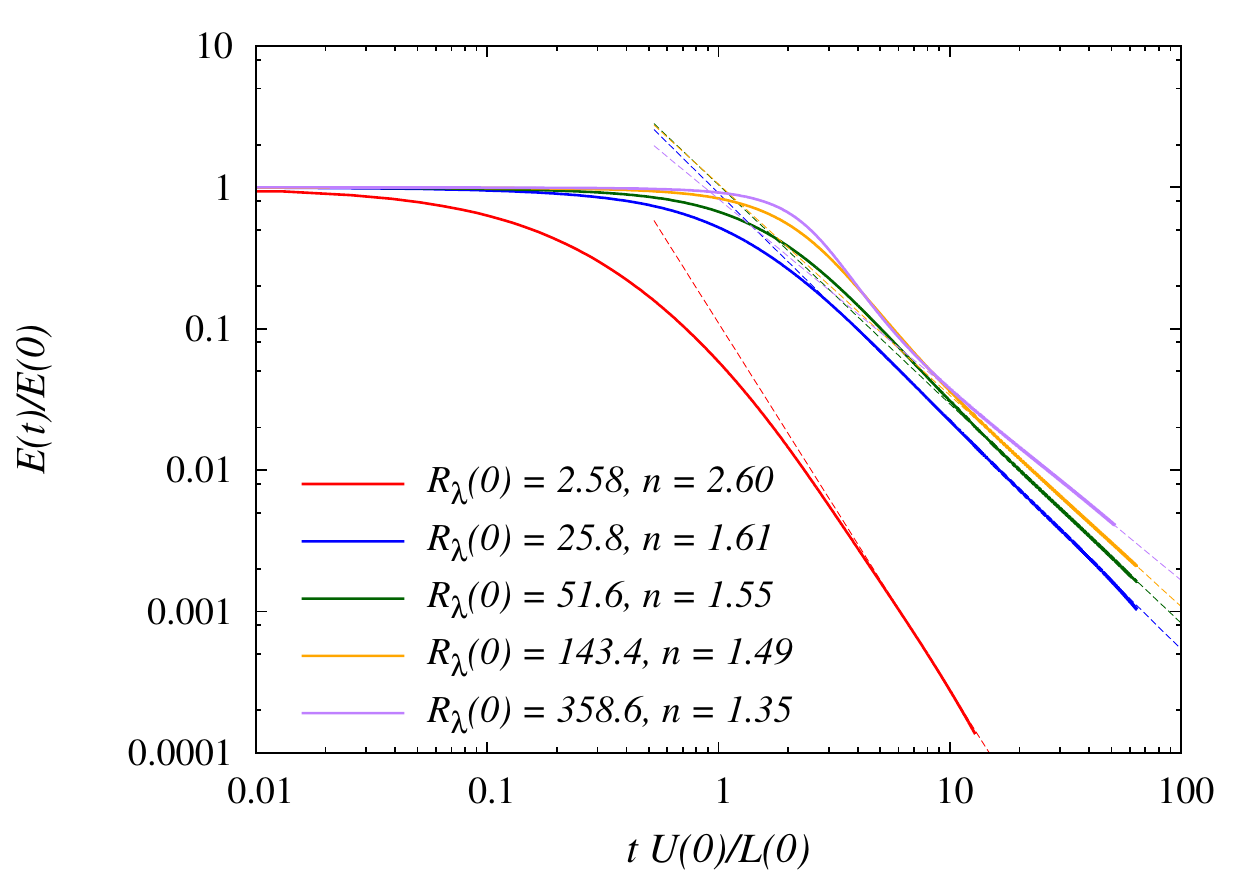}
 \end{center}
 \caption{Identification of power-law decay of the total energy. The total 
 energy is plotted against time on logscales, such that a period satisfying 
 power-law decay can be identified by a straight line.}
 \label{fig:powerlaw_decay}
\end{figure}

We begin with the direct determination of power-law decay and then move
on to the indirect method based on the Taylor microscale.
The turbulent kinetic energy and dissipation rate decay are generally
found to decay with time as
  \begin{equation}
   E(t) \propto t^{-n} \ , \qquad \varepsilon(t) \propto t^{-n-1} \ .
  \end{equation}
In Figure \ref{fig:powerlaw_decay}, we show the decay curves for the
energy $E(t)$ divided by its initial value, for a range of initial
values of the Reynolds number, ranging over $2.58 \leq \Rl \leq 358.6$.
The figure also shows the measured exponents for the decay and these were
found to decrease with increasing Reynolds number. Table
\ref{tbl:exponent_summary} summarises the experimental and theoretical
situation regarding decay exponents. It can be seen from there that
our values for exponents are not out of line with the field in general.
In view of the lack of consensus in this field, we can make no
stronger statement than that.

\begin{figure}[tbp]
 \begin{center}
  \includegraphics[width=0.6\textwidth]{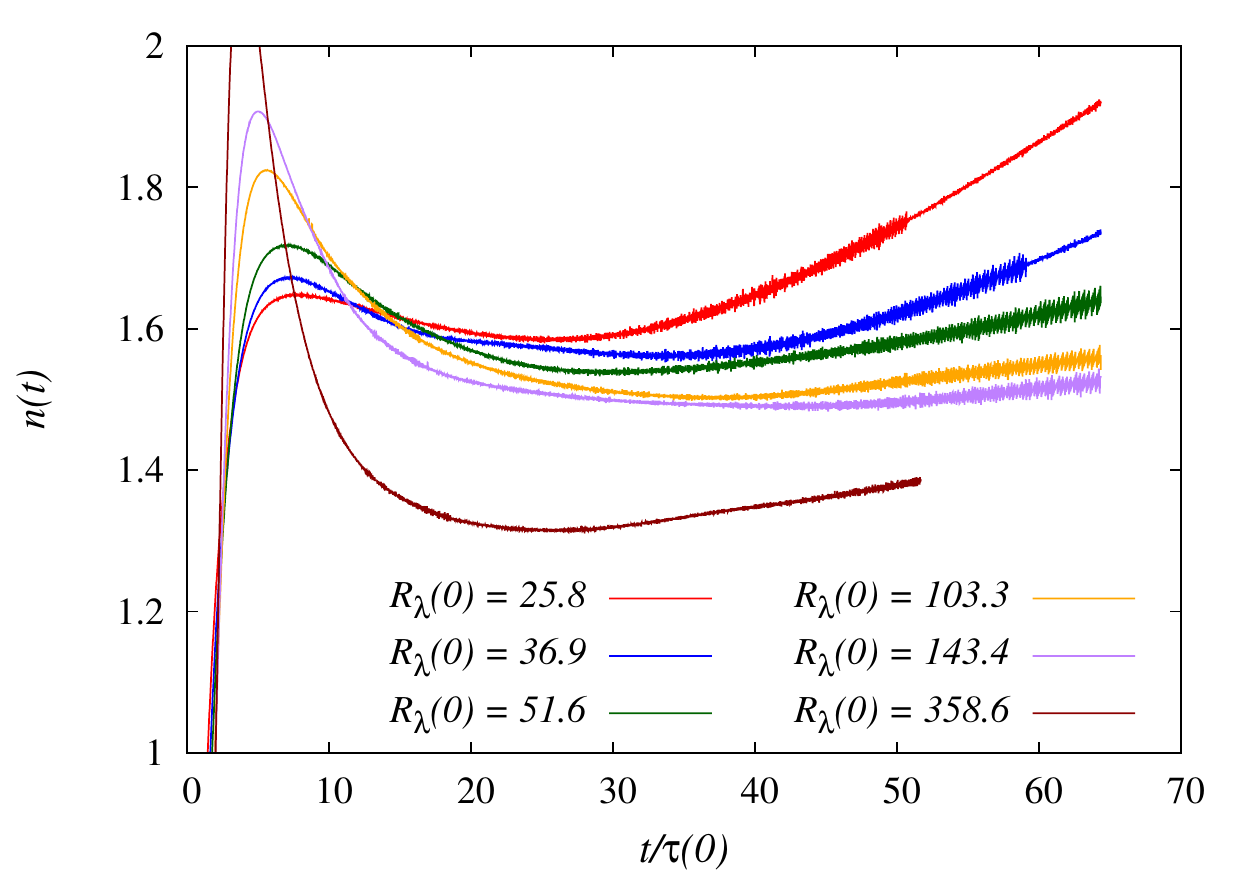}
 \end{center}
 \caption{The local slope for the decay of the total energy, plotted against 
 (scaled) time. A plateau indicates a region of power-law decay.}
 \label{fig:measure_decay_exp}
\end{figure}

In Fig. \ref{fig:measure_decay_exp} we illustrate the way in which we
determined the decay exponent by measuring the local slope of the decay
curves. This is based on the procedure
  \begin{equation}
   -n(t) = \frac{d\log{E}}{d\log{t}} = \frac{t}{E} \frac{dE}{dt} \ .
  \end{equation}
Evidently when we plot $n(t)$ against time on linear scales, as in Fig.
\ref{fig:measure_decay_exp}, a plateau corresponds to a region of
power-law decay.

\subsection{The onset of $t^{1/2}$ scaling of $\lambda$}

As an alternative criterion, we can make use of this well known exact
result for the decay of the Taylor microscale $\lambda$ in isotropic
decaying turbulence, as referred to in Section 3.1.

\begin{figure}[tbp]
 \begin{center}
  \includegraphics[width=0.6\textwidth]{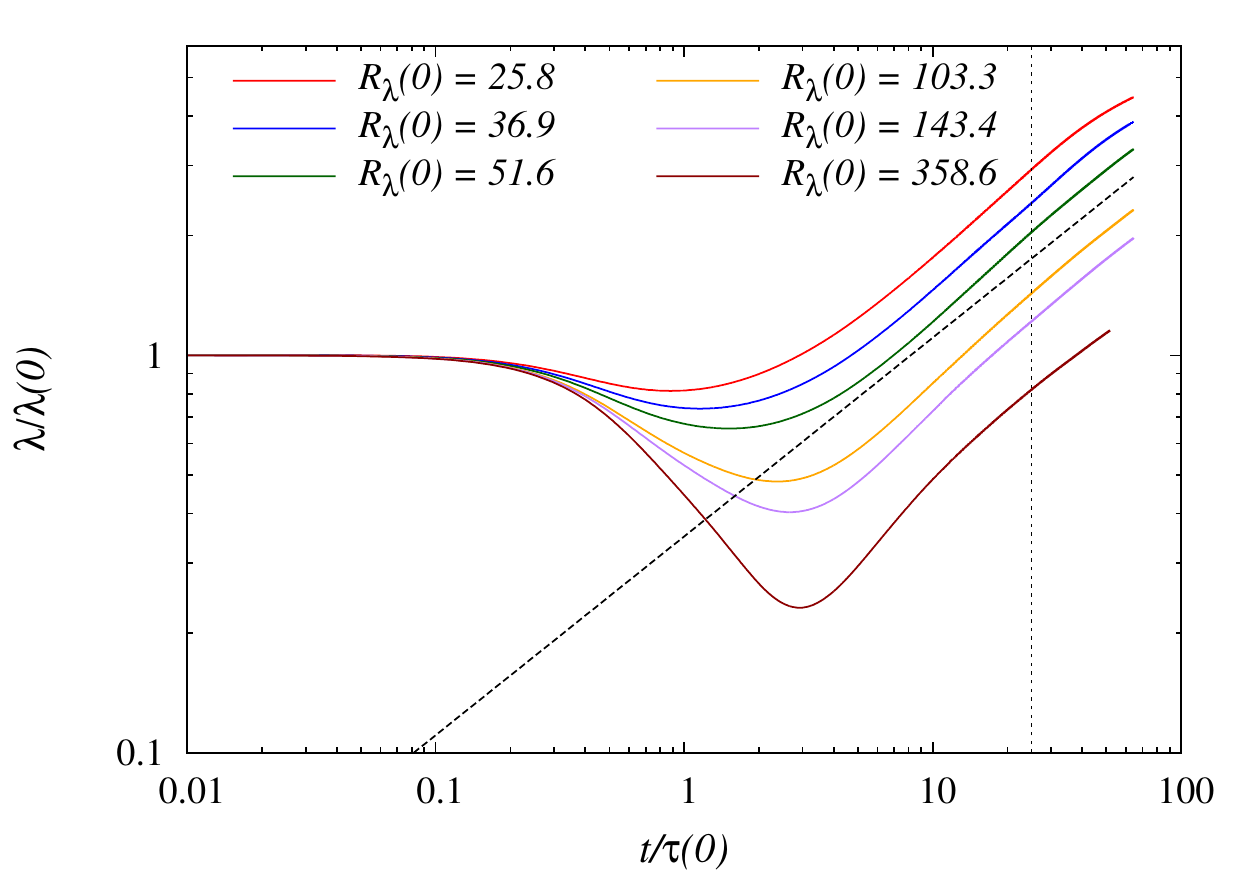}
 \end{center}
 \caption{Identification of $t^{1/2}$ behaviour of the Taylor microscale, 
 $\lambda$, for several different Reynolds numbers. The vertical dotted line 
 indicates the time after which we appear to observe $\sqrt{t}$ scaling, as
 shown by the dashed line.}
 \label{fig:lambda_decay}
\end{figure}

Using standard relationships for decay and dissipation rates in
isotropic turbulence
(e.g. see \cite{McComb14a})
we may write: \begin{equation}
   \frac{d E}{dt} = \frac{3}{2} \frac{d U^2}{dt} = -\varepsilon = - 
   \frac{15\nu_0 U^2}{\lambda^2}
  \end{equation}
Then, for power-law decay, where $U^2 \propto t^{-n}$, we have
  \begin{equation}
   \label{eq:powerlaw_decay}
   -nt^{-n-1} = -\frac{10\nu_0}{\lambda^2} t^{-n} \qquad\Longrightarrow 
   \qquad \lambda^2 = \frac{10\nu_0}{n} t \qquad\Longrightarrow \qquad 
   \lambda \propto \sqrt{t} \ ,
  \end{equation}
for all exponents $n$. We note that Figure \ref{fig:lambda_decay} shows
$\sqrt{t}$-scaling for $t\gtrsim 25\tau(0)$, where $\tau(0) =
L(0)/U(0)$. Clearly this rules out exponential decay because, if $U^2
\propto \exp(-\alpha t)$, then we have the well known result
\begin{equation}
   -\alpha\exp(-\alpha t) = -\frac{10\nu_0}{\lambda^2} \exp(-\alpha t) 
   \qquad\Longrightarrow\qquad \lambda^2 = \frac{10\nu_0}{\alpha} \qquad 
   \Longrightarrow \qquad \lambda = \text{constant} \ .
\end{equation}

The idea of using the behaviour of the Taylor microscale as a way of
obtaining the decay exponent for the energy has been explored by various
workers in the field. For example, Huang and Leonard \cite{Huang94},
Burattini \emph{et al} \cite{Burattini06}, and Lavoie \emph{et al}
\cite{Lavoie07}. In the interests of completeness, we present some
results of our own for this procedure in Appendix B. However, we note
here that the investigation of Burattini \emph{et al} \cite{Burattini06}
is of particular interest to us as they found that their results for
$C_\vep$ in free decay were comparable to those for forced turbulence.

\section{Energy decay exponent from the decay of the Taylor microscale}

As mentioned in Appendix A, we can obtain a value of the exponent $n$
from the decay of the Taylor microscale. We do this by rearranging the
relationship for $\lambda^2(t)$, as given in equation
(\ref{eq:powerlaw_decay}), and by introducing a virtual origin $t_0$. In
practice, uncertainty about the value of the virtual origin requires
iterative methods. We will retain it in our formulation but for sake of
simplicity we will set it equal to zero in calculations.

\begin{figure}[tbp]
 \begin{center}
  \includegraphics[width=0.6\textwidth]{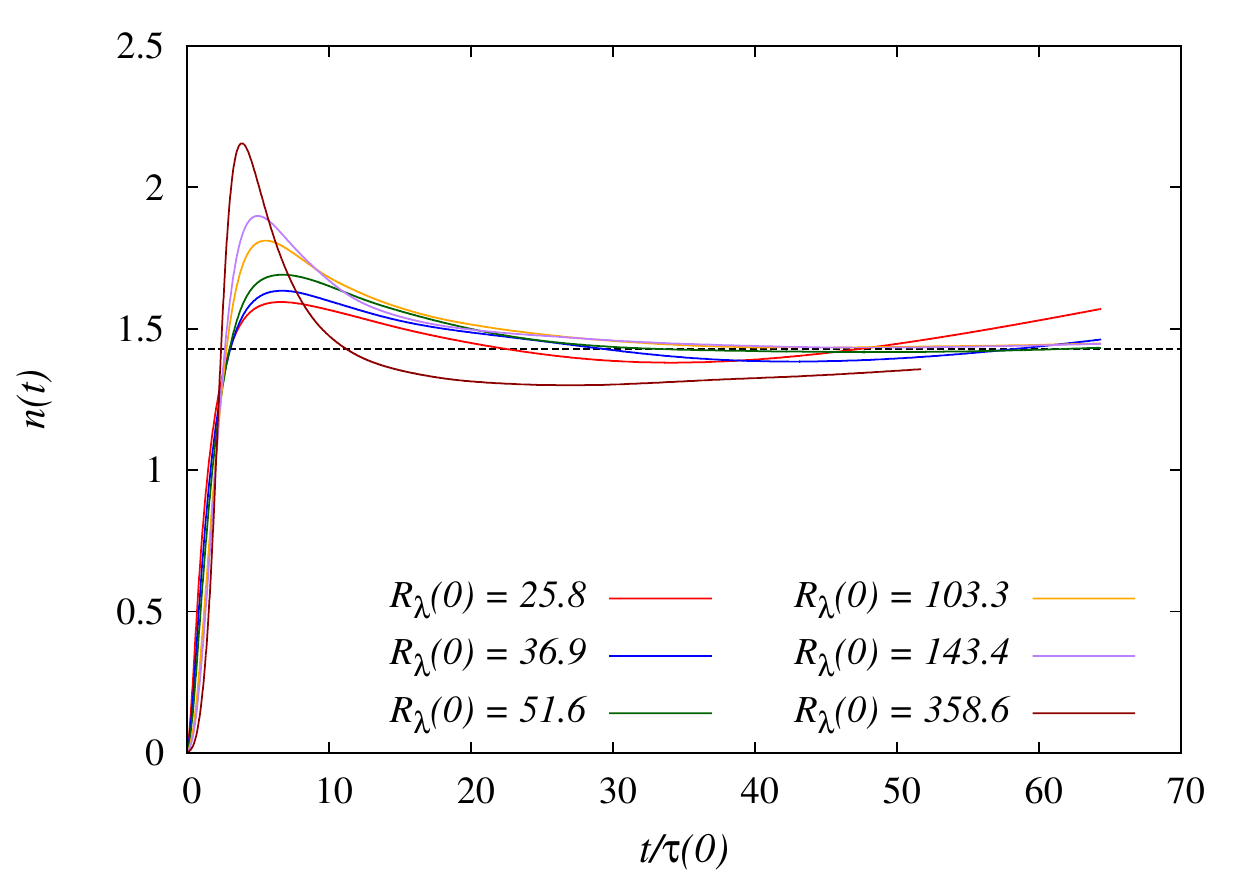}
 \end{center}
 \caption{Evaluation of the energy decay exponent using the Taylor 
 microscale, $n = 10\nu(t-t_0)/\lambda^2(t)$ with $t_0 = 0$. The horizontal 
 line indicates the Kolmogorov value of $n = 10/7$.}
 \label{fig:decay_exp_lambda}
\end{figure}

We find that
  \begin{equation}
   n(t) = \frac{10\nu_0}{\lambda^2(t)} (t - t_0) \ ,
  \end{equation}
where $t_0$ is the virtual origin of the power-law decay. Figure
\ref{fig:decay_exp_lambda} shows $n(t)$ calculated in this manner with
$t_0 = 0$. The dashed horizontal line indicates the Kolmogorov value of
$10/7$. This is a simplified version of what is considered later, in
Figure \ref{fig:decay_exp}.

\begin{figure}[tbp]
 \begin{center}
  \includegraphics[width=0.6\textwidth]{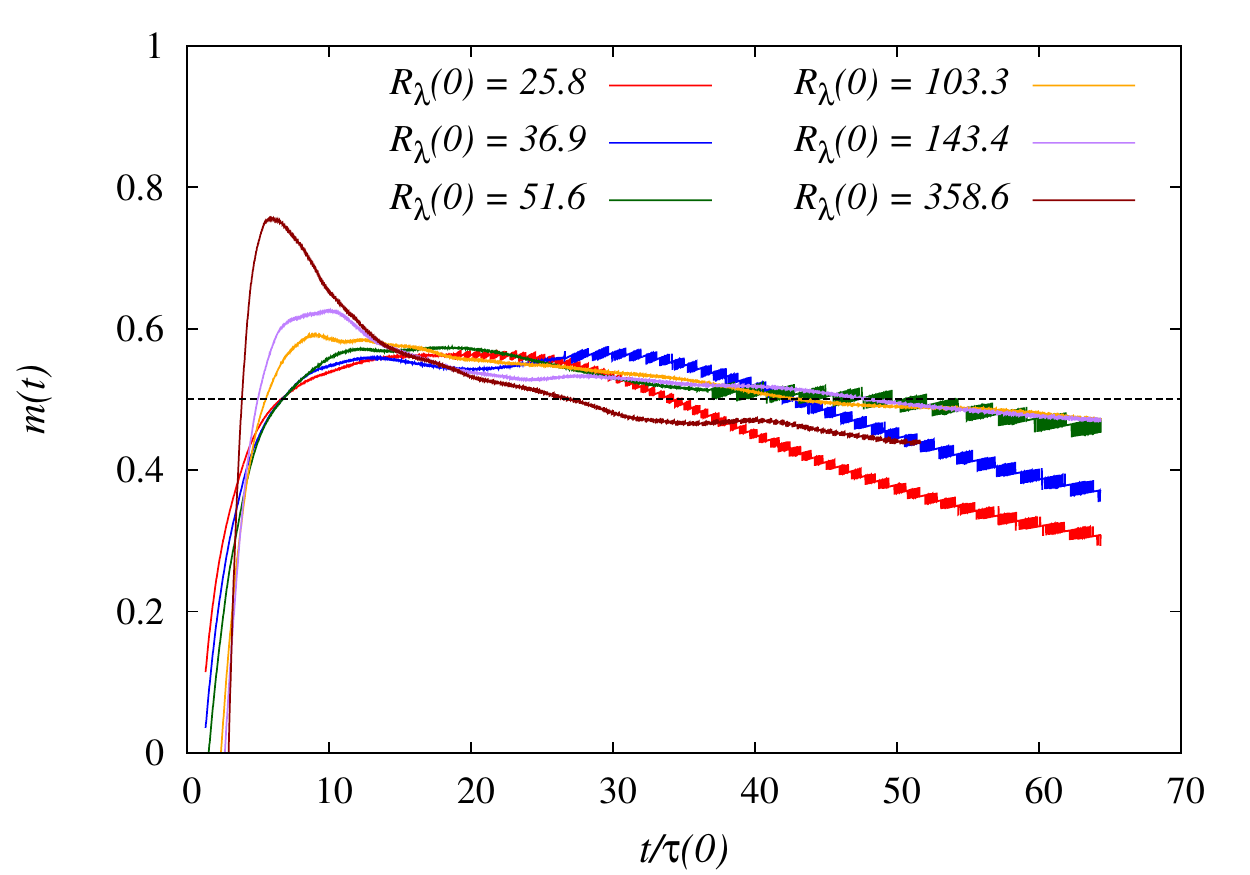}
 \end{center}
 \caption{The local slope for the time-variation of the Taylor microscale, 
 $\lambda$, such that $\lambda(t) \propto t^{m}$. A plateau at 0.5, as 
 indicated by the horizontal dashed line, would represent a region of 
 power-law decay for the total energy.}
 \label{fig:lambda_decay_exp}
\end{figure}

\begin{figure}[tbp]
 \begin{center}
  \includegraphics[width=0.45\textwidth]{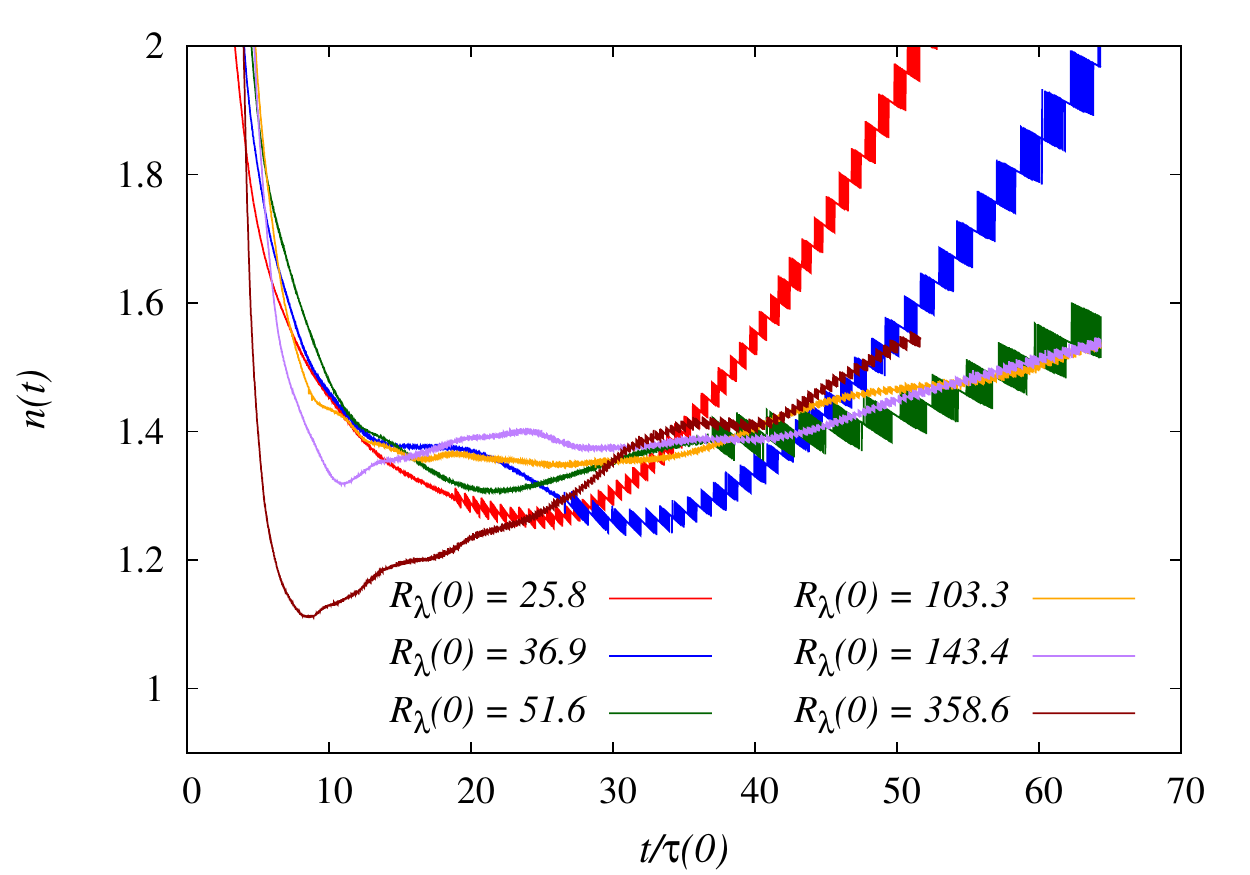}
  \includegraphics[width=0.45\textwidth]{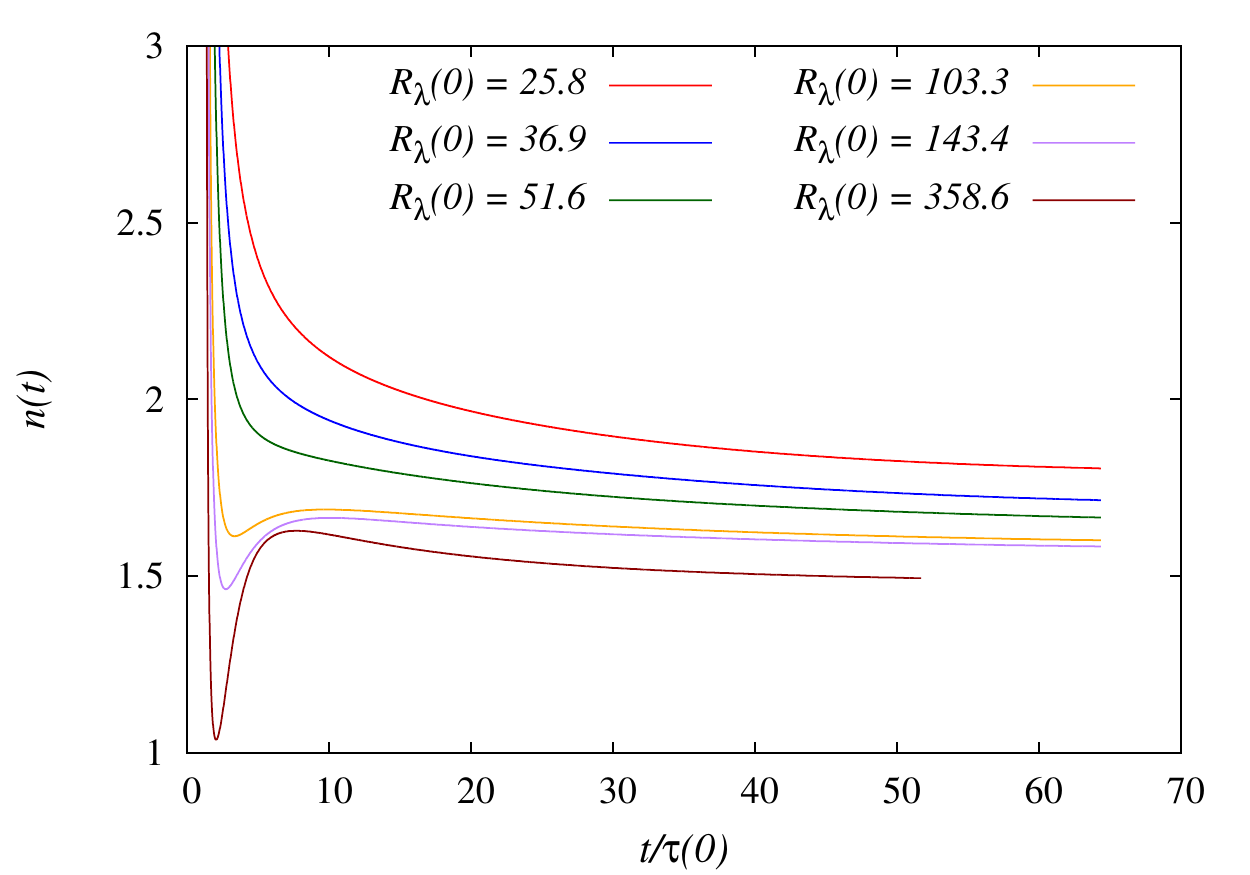}
 \end{center}
 \caption{Calculation of the power-law decay exponent for the total energy.  
 (Left) Calculated from $d\lambda^2/dt$. (Right) Calculated using 
 $(1/\log{t})\int^t_0 ds\ 1/\lambda^2$.}
 \label{fig:decay_exp}
\end{figure}

In order to find the decay exponent $m$ of the microscale, we proceed much
as we did for the decay of the energy. Consider $\lambda(t) \sim t^{m}$,
where $m = 1/2$ corresponds to power-law decay of the total energy. As
before, we measure the local slope,
\begin{equation}
   m(t) = \frac{d\log{\lambda}}{d\log{t}} = \frac{t}{\lambda} 
   \frac{d\lambda}{dt} \ .
\end{equation}
In Figure \ref{fig:lambda_decay_exp}, we plot the resulting $m(t)$
against time. A plateau at 1/2, shown by the horizontal dashed line on
the graph, would indicate a region of power-law decay of the total energy.

We can now obtain the energy exponent $n$ from the behaviour of
$\lambda$. To do this, we take a derivative with respect to $t$ of
equation \eqref{eq:powerlaw_decay}, to obtain:
\begin{equation}
   \frac{d \lambda^2(t)}{dt} = \frac{10\nu_0}{n} \ ,
  \end{equation}
from which we deduce that
\begin{equation}
   n = 10\nu_0\bigg/\frac{d \lambda^2(t)}{dt} \ .
  \end{equation}
Note that this form is not dependent on the time when power-law decay 
starts, $t_0$. This is plotted in the left-hand figure of Figure
\ref{fig:decay_exp}. 


Alternatively, we could take $U^2 \sim (t-t_0)^{-n(t)}$, in which case
for equation \eqref{eq:powerlaw_decay} we have
\begin{equation}
   \frac{\partial}{\partial t} \Big( n(t) \log(t-t_0) \Big) = 
   \frac{10\nu_0}{\lambda^2(t)} \ ,
  \end{equation}
which we integrate (with $n(0) = 0$) to find
\begin{equation}
   n(t) = \frac{10\nu_0}{\log(t-t_0)} \int_0^t \frac{ds}{\lambda^2(s)} \ .
  \end{equation}
This is clearly dependent on $t_0$. Note: Fukayama \etal\
\cite{Fukayama00} comment on numerical  integration being more stable
than numerical differentiation. The results of this procedure are
presented  in the right-hand figure of Figure \ref{fig:decay_exp}, with $t_0 =
0$.  Increasing $t_0$ would have the  effect of lowering the curves.

\end{document}